\documentclass[preprint,nofootinbib,superscriptaddress,amsmath,amssymb,UTF8]{revtex4}
\usepackage{amsmath,mathrsfs,amsbsy,color,graphicx,bm,amsthm,amsfonts}
\usepackage{bbm}
\usepackage{times}
\usepackage{units}
\usepackage{dcolumn}
\usepackage{graphicx}
\usepackage{epsfig}
\usepackage{epstopdf}
\usepackage[colorlinks,linkcolor=red,anchorcolor=green,citecolor=black,CJKbookmarks=True]{hyperref}
\DeclareMathSymbol{\shortminus}{\mathbin}{AMSa}{"39}
\usepackage{mathrsfs}
\usepackage{braket}
\usepackage{amssymb}
\usepackage{txfonts}
\usepackage{float}
\usepackage{enumitem}
\usepackage{multirow}
\usepackage{orcidlink}
\newcommand{\udots}{\mathinner{\mskip1mu\raise1pt\vbox{\kern7pt\hbox{.}}
\mskip2mu\raise4pt\hbox{.}\mskip2mu\raise7pt\hbox{.}\mskip1mu}}
\begin{document}
\title{Influence of dark matter on quantum entanglement and coherence in curved spacetime}

\author{Shu-Min Wu\orcidlink{0000-0001-7392-1327}}
\email[]{smwu@lnnu.edu.cn} 
\affiliation{Department of Physics, Liaoning Normal University, Dalian 116029, China}

\author{Yu-Xuan Wang}
\affiliation{Department of Physics, Liaoning Normal University, Dalian 116029, China}

\author{Si-Han Shang}
\affiliation{Department of Physics, Liaoning Normal University, Dalian 116029, China}

\author{Wentao Liu\orcidlink{0009-0008-9257-8155}}
\email[]{wentaoliu@hunnu.edu.cn (corresponding author)} 
\affiliation{Lanzhou Center for Theoretical Physics, Key Laboratory of
 Theoretical Physics of Gansu Province, Key Laboratory of Quantum Theory and Applications of MoE, Gansu Provincial Research Center for Basic Disciplines of Quantum Physics, Lanzhou University, Lanzhou 730000, China}
\affiliation{Institute of Theoretical Physics $ \& $ Research Center of Gravitation, 
School of Physical Science and Technology, Lanzhou University, Lanzhou 730000, China}


\begin{abstract}
Dark matter (DM) has not yet been directly detected, and phenomenological models such as perfect fluid dark matter (PFDM) provide analytically tractable settings for investigating its possible gravitational effects. Here PFDM denotes the conventional name of an anisotropic effective source, rather than a perfect fluid in the usual isotropic sense. In this work, we study the entanglement and coherence of bosonic and fermionic test fields near a Schwarzschild black hole surrounded by PFDM. These quantum fields probe the background geometry and are not assumed to be microscopic constituents of PFDM. Under our sign convention, only the $\alpha<0$ branch has a positive energy density and carries a physical PFDM interpretation, whereas the $\alpha>0$ branch is retained solely as a formal extension of the same metric family. Within the positive-density PFDM sector, entanglement and coherence vary monotonically with the PFDM parameter, while the complete extended parameter range exhibits a nonmonotonic dependence arising from the corresponding variation of the Hawking temperature. Bosonic entanglement is more susceptible to this geometric effect than fermionic entanglement, whereas fermionic coherence shows a stronger response than its bosonic counterpart. These results characterize how particle statistics and the choice of quantum resource govern the sensitivity of test fields to PFDM-induced modifications of the spacetime geometry.
\end{abstract}

\vspace*{0.2cm}
\maketitle
\section{introduction}
Quantum entanglement, a defining feature of multipartite quantum systems, arises from the complex tensor product structure of Hilbert spaces and the fundamental principle of quantum superposition.  It serves as a cornerstone of quantum information theory and a key resource enabling a wide range of revolutionary applications, including quantum teleportation, remote control, secure communication, and cryptography {\color{black}\cite{L1,L2,L3,L4,L5,L6,L7,L8}}. Over the past decades, significant advances have been made in characterizing various aspects of entanglement, such as its sudden death and revival, as well as its degradation or enhancement under different physical conditions {\color{black}\cite{L9,L10}}.
In parallel, quantum coherence, a broader concept rooted in the superposition of quantum states, is essential for the fundamental phenomena of quantum interference {\color{black}\cite{L11}}. Like entanglement, coherence is not only a pivotal resource in quantum technologies but also plays a crucial role in optical experiments and biological systems {\color{black}\cite{L12,L13,L14,L15,L16,L17}}. Importantly, the relationship between coherence and entanglement is both fundamental and intricate: coherence is a necessary condition for the emergence of entanglement, and in certain contexts, nonlocal coherence itself can be interpreted as a form of quantum entanglement {\color{black}\cite{L18,L19,L20}}. Despite the substantial progress made in understanding their interplay, many foundational questions regarding the connection and resource conversion between quantum coherence and entanglement remain open.

Relativistic quantum information combines quantum information, quantum field theory, and gravity. One research direction employs quantum networks and parameter estimation to probe spacetime structure {\color{black}\cite{L21,L22,L23,L24,L33}}, while quantum-enhanced measurements and entangled probes have also been studied in connection with gravitational waves {\color{black}\cite{L25,L26,L27,L28,L29,L30}}.
A complementary direction examines how spacetime geometry and observer motion affect quantum resources. Studies of noninertial frames address Unruh-induced entanglement, coherence, and quantum-channel properties {\color{black}\cite{L31,L32,L40,L43,L45,L50,L52,L53,LL70}}. In black-hole backgrounds, Hawking-induced changes in entanglement, nonlocality, and teleportation have been investigated {\color{black}\cite{L34,L35,L36,L38,L47,L48,L49,Wu:2025euf}}, alongside coherence, metrological quantities, and uncertainty relations {\color{black}\cite{L37,L39,L51,L60,L61,Li:2025jlu}}. Detector-based studies examine the roles of acceleration, boundaries, and vacuum fluctuations {\color{black}\cite{L62,L63,L69,Liu:2015oat,Liu:2025bzv}}, while other analyses consider coherence, quantum batteries, and correlations in cosmological and curved backgrounds {\color{black}\cite{L46,L54,L56,L57,L58,L64,L65,L66}}. Quantum-superposed geometries provide further settings for investigating entanglement and correlation harvesting {\color{black}\cite{L59,L70,Tang:2025mtc}}. Related work explores modified gravity, Lorentz violation, and their effects on acceleration radiation and quantum correlations {\color{black}\cite{L41,L42,L44,L55,AraujoFilho:2025nmc,Liu:2025hcx}}, as well as quantum-field simulation {\color{black}\cite{L67}}, black-hole light rings and shadows {\color{black}\cite{L68}}, and wormhole-induced correlations {\color{black}\cite{Liu:2025zts}}.

In contrast, the impact of DM on these quantum features remains largely underexplored in curved spacetime. DM, which constitutes about 27\% of the universe, does not interact electromagnetically and thus eludes direct detection, though its existence is strongly supported by indirect observational evidence {\color{black}\cite{L71,L72}}. Phenomenological descriptions include dark-matter halo profiles {\color{black}\cite{L73,L74}}, self-interacting, fuzzy, and superfluid dark matter {\color{black}\cite{L75,L76,L77}}, and primordial-black-hole candidates {\color{black}\cite{L78,L79}}. Dark-matter distributions have also been explored in fuzzy wormhole models {\color{black}\cite{ADO2,ADO6}} and Hayward black-hole configurations {\color{black}\cite{ADO4}}. Related studies of effective matter sources address static wormhole configurations {\color{black}\cite{ADO3}}, holographic-dark-energy-supported wormholes {\color{black}\cite{ADO5}}, and bulk-viscous cosmology {\color{black}\cite{ADO1}}.
Among these phenomenological approaches, PFDM provides an analytically tractable effective-fluid geometry and has been studied in connection with galactic rotation curves {\color{black}\cite{L80,L81}}, rotating black holes and shadows {\color{black}\cite{L88,L89,L82}}, thermodynamics {\color{black}\cite{L83}}, particle motion {\color{black}\cite{L84}}, accretion-disk dynamics {\color{black}\cite{L85}}, and quasinormal modes {\color{black}\cite{L86}}. In the present work, PFDM is used as a phenomenological gravitational background rather than as a microscopic particle model. Our purpose is therefore to characterize how quantum resources carried by test fields respond to PFDM-induced modifications of the spacetime geometry, rather than to propose a direct experimental scheme for detecting dark-matter particles.

Motivated by these considerations, we investigate the entanglement and coherence of fermionic and bosonic test fields in the spacetime of a Schwarzschild black hole surrounded by PFDM. Alice and Bob are idealized observers used to define access to different quantum-field modes: Alice is described as a Kruskal observer in the asymptotically flat region, whereas Bob has access only to the exterior modes outside the event horizon. This observer-mode construction follows the widely used framework of relativistic quantum information {\color{black}\cite{L31,L32,L34,L38}} and does not prescribe a specific detector deployment near an astrophysical black hole. We derive the corresponding quantum-information measures and compare their responses to the effective-fluid-induced geometric deformation. We distinguish throughout between the positive-density sector $\alpha<0$, to which the physical PFDM interpretation is restricted, and the formal extension $\alpha>0$, to which no conventional dark-matter interpretation is assigned. This distinction shows that the physical PFDM sector produces a monotonic response, whereas the recovery responsible for the nonmonotonic behavior over the complete parameter range occurs in the formally extended branch. The different sensitivities of bosonic entanglement and fermionic coherence thereby characterize field-statistics-dependent quantum signatures of the modified geometry.

The paper is organized as follows. In Sec. II, we present a systematic derivation of the quantization procedures for both bosonic and fermionic fields in the background of a Schwarzschild black hole surrounded by a PFDM halo.  Sec. III introduces the measures employed to quantify quantum coherence and entanglement. In Sec. IV, we investigate the behavior of these quantum resources for both types of fields in the PFDM-modified Schwarzschild background. The last section is devoted to a brief conclusion.  Throughout this paper, we adopt the natural units
$G=\hbar=c=k_B=1$, unless otherwise specified.

\section{ Quantization of both fermionic and bosonic fields  in the  Schwarzschild black hole surrounded by PFDM halo }\label{sec2}
The Einstein field equations represent the relationship between spacetime curvature and the distribution of matter and energy as
\begin{equation}\label{GEQ}
R_{\mu\nu}-\frac{1}{2}g_{\mu\nu}R=8\pi T_{\mu\nu},
\end{equation}
where $ R_{\mu\nu} $ is the Ricci tensor, characterizing the local curvature of spacetime due to matter, and $ R $ is the Ricci scalar, providing a measure of the overall curvature of spacetime.
$ T_{\mu\nu} $ is the energy-momentum tensor, which represents the distribution of energy, momentum, and stress (such as pressure and shear) within spacetime. 
It is well known that the field equation (\ref{GEQ}) admits a simple vacuum solution when $ T_{\mu\nu}=0 $, which describes a Schwarzschild black hole spacetime.
The study of information propagation near the event horizon of this black hole was a central focus of past research in the field of relativistic quantum information {\color{black}\cite{L34,L35,L36,L37,L38}}.

Astrophysical black holes need not be isolated, and one widely studied phenomenological description of their environment is the model conventionally termed perfect fluid dark matter (PFDM). We use this name only to identify the metric family studied in the literature; the effective source is anisotropic and is not a perfect fluid in the standard sense. In this model, the effective source entering equation (\ref{GEQ}) is written as $T^{\mu}_{~\nu}=\text{diag}[-\rho,p_r,p_\theta,p_\varphi]$, where $T^{\mu}_{~\nu}=g^{\mu\sigma}T_{\sigma\nu}$. Its energy density, radial pressure, and tangential pressures are
\begin{align}\label{eq222}
\rho=-p_r=-\frac{\alpha}{8\pi r^3}, \quad p_\theta=p_\varphi=-\frac{\alpha}{16\pi r^3}.
\end{align}
Here $\alpha$ is the dimensional PFDM parameter, which has dimensions of length, and $\tilde{\alpha}\equiv\alpha/M$ is its dimensionless counterpart. Thus, the notation $\alpha/M$ used on the axes of the figures is exactly equivalent to $\tilde{\alpha}$. Throughout this paper, $\alpha$ denotes the dimensional parameter, whereas $\tilde{\alpha}$ or, equivalently, $\alpha/M$ denotes its dimensionless counterpart. With the convention in equation (\ref{eq222}), $\alpha<0$ (equivalently, $\tilde{\alpha}<0$) corresponds to a positive energy density, whereas $\alpha>0$ (equivalently, $\tilde{\alpha}>0$) corresponds to a negative energy density.
A perfect fluid has equal principal pressures in its local rest frame. In contrast, equation (\ref{eq222}) gives $p_r=-\rho$ and $p_t\equiv p_\theta=p_\varphi=\rho/2$. Following the average-pressure definition used for Kiselev-type sources in Ref.~{\color{black}\cite{VisserPF}}, we define
\[
\bar p\equiv\frac{p_r+2p_t}{3},\qquad
w_{\mathrm{eff}}\equiv\frac{\bar p}{\rho}=0\quad(\alpha\ne0).
\]
This value of $w_{\mathrm{eff}}$ describes a vanishing spatial trace of the pressure tensor, not pressureless matter: $p_r/\rho=-1$, $p_t/\rho=1/2$, and $p_t-p_r=3\rho/2$ remain nonzero. At $\alpha=0$ the effective source vanishes, so the ratio $\bar p/\rho$ is undefined and the geometry is Schwarzschild. The source in equation (\ref{eq222}) is a phenomenological ansatz, not a microscopic model of isotropic cold dark matter.

For the chosen source, it is convenient to write Einstein's equations in residual form,
\begin{align}\label{GEQ1}
\mathcal{E}^{\mu}_{~\nu}\equiv R^\mu_{~\nu}-\frac{1}{2}\delta^\mu_{~\nu}R-8\pi T^\mu_{~\nu}=0.
\end{align}
We adopt the static, spherically symmetric single-function ansatz in curvature coordinates $ (t,r,\theta,\varphi) $,
\begin{equation}\label{ds22}
\begin{aligned}
ds^2=&-F(r)dt^2+F(r)^{-1}dr^2+r^2 d\theta^2+r^2\sin^2\theta d\varphi^2.
\end{aligned}
\end{equation}
The restriction $g_{tt}g_{rr}=-1$ makes the temporal and radial mixed components of the Einstein tensor equal and hence requires $p_r=-\rho$. The additional constitutive choice $p_t=\rho/2$, or equivalently $\bar p=0$, then fixes the density profile through radial stress-energy conservation:
\[
p_r'+\frac{F'}{2F}(\rho+p_r)+\frac{2}{r}(p_r-p_t)=0
\quad\Longrightarrow\quad
\rho'=-\frac{3\rho}{r}.
\]
Thus $\rho=C/r^3$, with $C=-\alpha/(8\pi)$ in our convention. This derivation explains the consistency of the ansatz with Einstein's equations; it does not establish this profile as a realistic galactic halo distribution. The nonzero components of the residual in equation (\ref{GEQ1}) are
\begin{equation}
\begin{aligned}
&\mathcal{E}^{t}_{~t}=\mathcal{E}^{r}_{~r}=\frac{F'(r)}{r}+\frac{F(r)-1}{r^2}-\frac{\alpha}{r^3},\\
&\mathcal{E}^{\theta}_{~\theta}=\mathcal{E}^{\varphi}_{~\varphi}=\frac{F''(r)}{2}+\frac{F'(r)}{r}+\frac{\alpha}{2r^3}.
\end{aligned}
\end{equation}
Solving $\mathcal{E}^{\mu}_{~\nu}=0$ yields the metric function
\begin{align}\label{eq:PFDMmetric}
F(r) = 1 - \frac{2M}{r} - \frac{\alpha}{r} \ln\left(\frac{|\alpha|}{r}\right),
\end{align}
where $\alpha$ characterizes the strength and sign of the effective-source contribution. This is the logarithmic branch displayed explicitly by Li and Yang {\color{black}\cite{L81}}: their relation $T^\theta_{~\theta}=(1-\epsilon)T^t_{~t}$ gives $p_t=\rho/2$ at $\epsilon=3/2$, and their solution agrees with equation (\ref{eq:PFDMmetric}) upon identifying $r_s=2M$ and $a=\alpha$.

The connection with the Kiselev construction requires distinguishing this logarithmic branch from the usual power-law expression. For the latter, $F_K(r)=1-2m_0/r-K/r^{1+3w}$, the nonvacuum source satisfies $p_r=-\rho$ and $p_t=(1+3w)\rho/2$, so $w=\bar p/\rho$ {\color{black}\cite{VisserPF}}. However, setting $w=0$ at fixed finite $K$ merely shifts the Schwarzschild mass and leaves no nonzero source outside the origin. Our metric instead has the mass function
\[
m(r)\equiv\frac r2[1-F(r)]
=M-\frac{\alpha}{2}\ln\!\left(\frac{r}{|\alpha|}\right).
\]
In the general mass-function formulation, Einstein's equations give $\rho=-p_r=m'/(4\pi r^2)$ and $p_t=-m''/(8\pi r)$. Hence $\bar p=0$ implies $(r m')'=0$, whose nonvacuum solution is logarithmic. We therefore use a generalized Kiselev-type anisotropic geometry with $w_{\mathrm{eff}}=0$, specifically the logarithmic branch, rather than the standard power-law Kiselev metric with $w$ simply set to zero. This specifies both the origin and the limited effective-source interpretation of the geometry used below.
In the continuous limit $\alpha\to0$, the metric reduces to the Schwarzschild solution.
In the following, the physical PFDM interpretation is restricted to the positive-density branch $\alpha<0$ (or $\tilde{\alpha}<0$). Results for $\alpha>0$ (or $\tilde{\alpha}>0$) are retained only as a formal extension of the same metric family for characterizing the response of quantum-information measures to the corresponding geometric deformation; this branch is not identified with astrophysical dark matter or dark energy.


\begin{figure}
\begin{minipage}[t]{0.5\linewidth}
\centering
\includegraphics[width=3.0in,height=5.2cm]{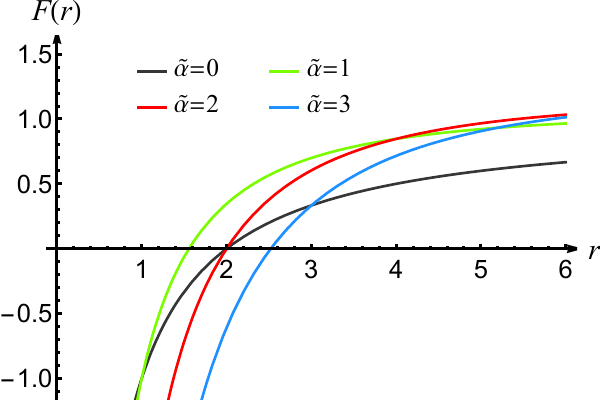}
\label{fig1a}
\end{minipage}%
\begin{minipage}[t]{0.5\linewidth}
\centering
\includegraphics[width=3.0in,height=5.2cm]{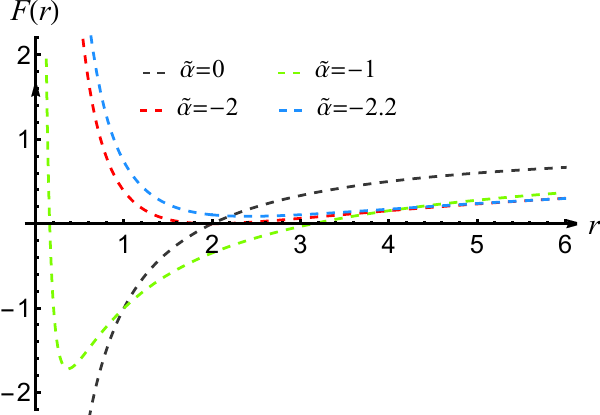}
\label{fig1b}
\end{minipage}%
\caption{The black-hole metric function $F(r)$ for various values of $\tilde{\alpha}$. The left panel shows the formally extended branch $\tilde{\alpha}\geq0$, whereas the right panel shows the positive-density PFDM sector $\tilde{\alpha}\leq0$.}
\label{Fig1}
\end{figure}

\begin{figure}
\includegraphics[scale=1.0]{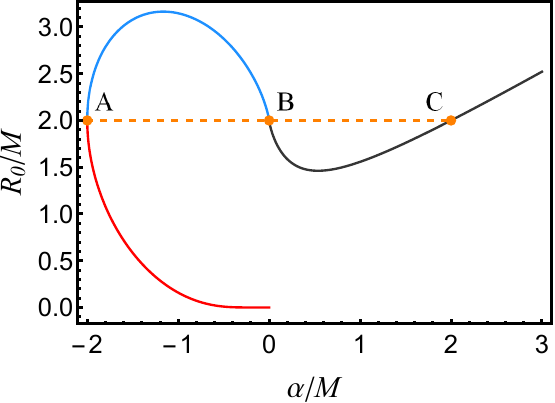}
\caption{The dependence of the black-hole coordinate singularity $R_0$ on the dimensionless parameter $\tilde{\alpha}$. The black curve represents the event horizon in the formally extended branch $\tilde{\alpha}>0$, the blue curve represents the event horizon in the positive-density PFDM sector $\tilde{\alpha}<0$, and the red curve depicts equation (\ref{rhh}) for $\tilde{\alpha}<0$.
 }
\label{Fig2}
\end{figure}

For the mathematical analysis of the metric family, the dimensional parameter $\alpha$ is treated as a free parameter. In terms of $\tilde{\alpha}=\alpha/M$, the existence of an event horizon requires $\tilde{\alpha}\geq-2$. Only the interval $-2\leq\tilde{\alpha}<0$ carries the positive-density PFDM interpretation, while $\tilde{\alpha}>0$ denotes the formal extension defined above.
Observational information on dark-matter pressure and density is important for assessing the physical interpretation of this effective source. Potapov et al.~{\color{black}\cite{Potapov2016}} used Galactic rotation-curve and mass information to assess an isotropic perfect-fluid halo and found a nearly pressureless description. Kumar et al.~{\color{black}\cite{Kumar2019}} constrained the dark-matter equation of state and sound speed using cosmic microwave background, baryon acoustic oscillation, and local Hubble-constant data, finding consistency with cold dark matter. These studies provide useful guidance for interpreting our model. The present source has anisotropic pressures, $p_r=-\rho$ and $p_t=\rho/2$, so its vanishing average pressure, $w_{\mathrm{eff}}=0$, should not be taken as evidence of agreement with these observational constraints. Applying the reported bounds requires accounting for the different source assumptions and physical scales. Accordingly, the interval $-2\leq\tilde{\alpha}<0$ is determined by positive density and horizon existence, rather than by an observational fit.

Within the assumed profile, a density normalization at a radius $r_0$ would determine $\alpha=-8\pi\rho(r_0)r_0^3$ in our units. Applying this relation to a specific system requires assessing the validity of the $r^{-3}$ density profile and the assumed stresses over the region of interest, particularly when connecting Galactic or cosmological information to the near-horizon environment. The parameter scans below therefore describe how quantum resources respond to the assumed geometry, while their astrophysical application requires observational validation of the background.

To visually represent this, we display the metric function $ F(r) $ for various $ \tilde{\alpha} $ in Fig.\ref{Fig1}.
Notably, when the parameter $ \tilde{\alpha} > 0 $, the location of the event horizon has an analytical expression,
\begin{align}\label{rhh}
r_h=\alpha W\left[ \frac{|\alpha|}{\alpha} \exp\left(\frac{2M}{\alpha}\right) \right].
\end{align}
Here, $ W\left[\cdot\right] $ is the Lambert $ W $ function.
When $ -2<\tilde{\alpha}\leq0 $, equation (\ref{rhh})  no longer corresponds to the event horizon of the black hole but instead represents the zero point to the left of the light cyan dashed line in the right panel of Fig.\ref{Fig1}.
This additional root does not enter our exterior-region quantum-information analysis, and we therefore do not assign it a further physical interpretation.
Although the analytical form of the event horizon cannot be obtained within the range $ -2<\tilde{\alpha}<0 $, we can numerically visualize the horizon position variation, as shown by the blue line segment between points A and B in Fig.\ref{Fig2}.
Interestingly, as shown in Fig. \ref{Fig2}, points A, B, and C represent the same event-horizon radius but correspond to three different values of the dimensionless parameter $\tilde{\alpha}$.
The blue–black curve corresponds to the outer horizon, with its blue (black) segment representing negative (positive) $\tilde{\alpha}$, while the red curve denotes the inner horizon.
In our analysis, only the outer horizon is physically relevant, since all near-horizon quantum information quantities are defined with respect to the exterior region of the event horizon {\color{black}\cite{L90}}.
For $\tilde{\alpha}=0$, the outer branch smoothly reduces to the Schwarzschild radius, and for negative $\tilde{\alpha}$ (blue segment) it continuously connects to this limit, remaining the physical solution.

We now analyze the surface gravity of black holes, which is commonly defined as the acceleration at the event horizon.
For a stationary particle, the four-velocity is expressed as
\begin{align}
u^\mu=\left\{u^0,0,0,0\right\},
\end{align}
where $ u^0 $ is determined by the normalization condition $ u^\mu u_\mu=-1 $.
The corresponding four-acceleration is given by
\begin{align}\label{amu}
a^\nu=u^\mu\nabla_\mu u^\nu=u^\mu\partial_\mu u^\nu+\Gamma^\nu_{\mu\rho}u^\mu u^\rho,
\end{align}
and the surface gravity, defined as the magnitude of the acceleration at the horizon, is
\begin{align}
\kappa=\left.\frac{1}{2}\frac{dF(r)}{dr}\right|_{r=r_h}=\frac{1}{2r_h^2}\left[2M+\alpha+\alpha\log\left(\frac{|\alpha|}{r_h}\right)\right],
\end{align}
By solving equation (\ref{rhh}), the above expression simplifies to
\begin{align}
\kappa=\frac{\alpha+r_h}{2r_h^2},
\end{align}
which is valid for both positive and negative values of the dimensional parameter $\alpha$.
In an arbitrary theory of gravity, accounting for the effects of quantum particle creation, the Hawking temperature of a black hole with constant surface gravity is given by $ T = \kappa / 2\pi $, as detailed in {\color{black}\cite{L91}}.
For $\alpha<0$, let $x=r/M$ and $\beta=-\tilde{\alpha}=-\alpha/M>0$. The horizon equation can then be written as $H_\beta(x)=x-2+\beta\ln(\beta/x)=0$. Since $H_\beta'(x)=1-\beta/x$, the minimum of $H_\beta(x)$ occurs at $x=\beta$, where $H_\beta(\beta)=\beta-2$. Therefore, for $0<\beta<2$, equivalently $-2<\tilde{\alpha}<0$, the outer horizon satisfies $x_h>\beta$, and hence $r_h+\alpha=M(x_h-\beta)>0$. It follows that $\kappa=(r_h+\alpha)/(2r_h^2)>0$. At the limiting value $\tilde{\alpha}=-2$, the two horizons coincide at $r_h=2M=-\alpha$ and $\kappa=0$. Thus, the condition $\tilde{\alpha}\geq-2$, together with the choice of the outer horizon, guarantees $\kappa\geq0$. For $\alpha>0$, the same conclusion follows immediately from $r_h>0$.
We next introduce massless scalar and Dirac fields as independent test quantum fields on this fixed background. They encode the quantum information shared by Alice and Bob and are not assumed to be microscopic constituents of PFDM. No direct nongravitational interaction between these test fields and the effective source is introduced; the source affects them only through the modified spacetime geometry.

\subsection{Klein-Gordon Equation  }\label{sec2.1}
The massless scalar field satisfies the Klein-Gordon equation given by
\begin{align}
\nabla_\mu\nabla^\mu\psi_s = \frac{1}{\sqrt{-g}}\partial_\mu\left(g^{\mu\nu}\sqrt{-g}\partial_\nu\psi_s\right) = 0,
\end{align}
where the scalar field $\psi_s$ can be expressed in the form
\begin{align}
\psi_s(t,r,\theta,\varphi) = \psi_s(t,r)Y^{lm}(\theta,\varphi),
\end{align}
and $Y^{lm}(\theta,\varphi)$ represents the scalar spherical harmonics.
By separating the angular components, the radial part of the Klein-Gordon equation becomes
\begin{align}
\left(F\partial_r^2 - F^{-1}\partial_t^2\right)\psi_s - \left(\frac{2F}{r} + F'\right)\partial_r\psi_s - \frac{l(l+1)}{r^2}\psi_s = 0,
\end{align}
where $l$, the azimuthal quantum number, appears as a separation constant.
Through appropriate coordinate and function transformations, the equation of motion for the scalar field $\psi_s$ can be rewritten in the form of a Schr\"{o}dinger-like wave equation.
To achieve this, we set $\psi_s(t,r) = e^{-i\omega_s t}r^{-1}Z_s(r)$, leading to the equation
\begin{align}\label{mastereq}
\left(\partial^2_{r_*} +\omega_s^2\right)Z_s = \mathcal{V}_sZ_s,
\end{align}
where $r_*$ represents the tortoise coordinate, defined as
\begin{align}
r_*=\int\frac{dr}{1-\frac{2M}{r}-\frac{\alpha}{r}\ln\left(\frac{|\alpha|}{r}\right)}.
\end{align}
The term $\mathcal{V}_s$ denotes the effective potential for the scalar field and is given by
\begin{align}
\mathcal{V}_s=F\left[\frac{l(l+1)}{r^2}+\frac{2M}{r^3}+\frac{\alpha+\alpha \ln\left(\frac{|\alpha|}{r}\right)}{r^3}\right].
\end{align}

\subsection{Dirac Equation }\label{sec2.2}

The massless Dirac field obeys the Dirac equation, expressed as
\begin{align}\label{dirac}
\gamma^ae_a^{~\mu}\left(\partial_\mu+\Gamma_\mu\right)\psi_d=0.
\end{align}
To explicitly derive the Dirac equation in the PFDM-surrounded Schwarzschild spacetime, we first introduce the tetrad formalism. The curved spacetime gamma matrices are related to the flat spacetime gamma matrices through
\[\gamma^\mu=e^\mu_a\gamma^a ,\]
where $e_a^{~\mu}$ is the inverse of the tetrad $e_\mu^{~a}$, which satisfies the relation
\begin{align}
g_{\mu\nu}=\eta_{ab}e_{\mu}^{~a}e_{\nu}^{~b}.
\end{align}
The explicit form of the tetrad inverse is given by
\begin{align}
e_a^{~\mu}=\left(\begin{array}{cccc}
1/\sqrt{F} & 0 & 0 & 0\\
0 & \sqrt{F} & 0 & 0 \\
0 & 0 & 1/r & 0 \\
0 & 0 & 0 & \csc\theta /r
\end{array}\right),
\end{align}
where  $ \eta_{ab}=\text{diag}\left(-1,1,1,1\right) $ is the Minkowski metric.
The Dirac matrices $\gamma^a$ are defined as follows: $\gamma^1 = i\tilde{\gamma}^3$, $\gamma^2 = i\tilde{\gamma}^1$, $\gamma^3 = i\tilde{\gamma}^2$, and $\gamma^0 = i\tilde{\gamma}^0$, where
\begin{align}
\tilde{\gamma}^0=\left(\begin{array}{cc}
O & I \\
I & O
\end{array}\right),~~~~~~~~
\tilde{\gamma}^i=\left(\begin{array}{cc}
O & \sigma_i \\
-\sigma_i & O
\end{array}\right),~~~~~ i=1,2,3.
\end{align}
Here, $I$ is the $2 \times 2$ identity matrix, $O$ is the $2 \times 2$ zero matrix, and $\sigma_i$ are the Pauli matrices. They obey the anti-commutation relation \[
\{\tilde{\gamma}^{a},\tilde{\gamma}^{b}\}=2\eta^{ab}.
\]
The spin connection $\Gamma_\mu$ is given by
\begin{align}
\Gamma_\mu=\frac{1}{8}\left[\gamma^a,\gamma^b\right]e_{a}^{~\nu}\left(\partial_\mu e_{b\nu}-\Gamma^\alpha_{\mu\nu}e_{b\alpha}\right),
\end{align}
where $\Gamma^\alpha_{\mu\nu}$ represents the Christoffel symbols, which are determined by the spacetime metric as
\[
\Gamma^\alpha_{\mu\nu}
=
\frac{1}{2}g^{\alpha\beta}
\left(
\partial_\mu g_{\beta\nu}
+\partial_\nu g_{\beta\mu}
-\partial_\beta g_{\mu\nu}
\right).
\]

Substituting into equation (\ref{dirac}), we obtain
\begin{equation}
\begin{aligned}
\left[\frac{\gamma^0}{\sqrt{F}}\partial_t+\gamma^1\sqrt{F}\left(\frac{1}{r}+\frac{F'}{4F}+\partial_r\right)\right.
\left.+\frac{\gamma^2}{r}\left(\frac{1}{2}\cot\theta+\partial_\theta\right)+\frac{\gamma^3\partial_\varphi}{r\sin\theta} \right]\psi_d=0,
\end{aligned}
\end{equation}
and using the Dirac field ansatz
\begin{align}\label{EQS}
\psi_d(t,r,\theta,\varphi)=\frac{e^{-i\omega_d t}e^{im\varphi}}{rF^{\frac{1}{4}}}
\left(\begin{array}{c}
\eta_-(r,\theta) \\ \eta_+(r,\theta)
\end{array}\right),
\end{align}
leads to the following equation
\begin{equation}
\left( \frac{\sigma_1}{2}\cot\theta +\sigma_1\partial_\theta+ \frac{im\sigma_2}{\sin\theta}+r\sqrt{F}\sigma_3\partial_r\mp\frac{i\omega_d r}{\sqrt{F}}\right)\eta_{\pm}(r,\theta)=0,
\end{equation}
where $\omega_d$ and $m$ denote the frequency and azimuthal quantum number of the Dirac field, respectively, with $m$ arising from the use of spherical harmonics in the field's multipole expansion.
Setting
\begin{align}
\eta_+(r,\theta)=\left(\begin{array}{c}
R_1(r)S_1(\theta) \\ R_2(r)S_2(\theta)
\end{array}\right),~~
\eta_-(r,\theta)=-\left(\begin{array}{c}
R_2(r)S_1(\theta) \\ R_1(r)S_2(\theta)
\end{array}\right),
\end{align}
leads to two sets of coupled first-order differential equations with separable variables
\begin{align}
\left(\frac{1}{2}\cot\theta-\frac{m}{\sin\theta}+\partial_\theta\right)S_1&=\lambda S_2,\\
\left(\frac{1}{2}\cot\theta+\frac{m}{\sin\theta}+\partial_\theta\right)S_2&=-\lambda S_1,
\end{align}
and
\begin{align}\label{R1}
\frac{d}{dr_*}R_1-i\omega_d R_1=\lambda\frac{\sqrt{F}}{r} R_2,\\ \label{R2}
\frac{d}{dr_*}R_2+i\omega_d R_2=\lambda\frac{\sqrt{F}}{r} R_1,
\end{align}
where $ \lambda=l+\frac{1}{2} $ is the separation constant, and the tortoise coordinate $ r_* $ is the same as defined in Sec. IIA.

In the field of relativistic quantum information, attention is primarily directed towards the radial equation {\color{black}\cite{L32,L34}}.
To obtain a Schr\"{o}dinger-like equation with a real effective potential for the radial equations (\ref{R1}) and (\ref{R2}), we employ the Chandrasekhar transformation as {\color{black}\cite{L92,L93}}:
\begin{align}
Z_d^{\pm}(r)=R_1(r)\pm R_2(r),
\end{align}
resulting in a set of decoupled Schr\"{o}dinger-like equations
\begin{align}\label{EQ1}
\left(\partial^2_{r_*} +\omega_d^2\right)Z^\pm_d(r_*) = \mathcal{V}^\pm_dZ^\pm_d(r_*),
\end{align}
where the effective potentials of Dirac fields are expressed as
\begin{equation}
\mathcal{V}_d^\pm=\frac{k_\pm^2F}{r^2}\mp \frac{k_\pm\sqrt{F}}{2r^2}\left[2r-6M-\alpha-3\alpha\ln\left(\frac{|\alpha|}{r}\right)\right],
\end{equation}
with \( k_\pm=l+\frac{1}{2}\pm\frac{1}{2} \).
Here, $l$ is the angular quantum number, taking values in the set \(\{...-\frac{3}{2},-\frac{1}{2},+\frac{1}{2},+\frac{3}{2}...\} \).

\subsection{ Vacuum Structure Near the Event Horizon}

Solving equations (\ref{EQS}) and (\ref{EQ1})  near the event horizon $ r_h $, we need to perform a Taylor expansion of the effective potentials at the horizon; the analytical solutions are
\begin{align}
Z_{s,d}(t,r_*)=C_1e^{-it\omega_{s,d}}e^{ir_*\omega_{s,d}}+C_2e^{-it\omega_{s,d}}e^{-ir_*\omega_{s,d}}.
\end{align}
Note that for fermions, we only consider the positive mode frequencies.
The complex coefficients $ C_1 $ and $ C_2 $ are determined by the boundary conditions, with $ C_1=0 $ corresponding to a pure ingoing wave and $ C_2=0 $ to a pure outgoing wave, respectively.
Upon second quantization of the fields, we extend the outgoing wave functions to cover both regions outside ($I$) and inside ($II$) the black hole event horizon, using the light-cone coordinates $ v=t+r_* $ and $ u=t-r_* $, given by
\begin{align}
Z^{I}_{s,d}(r>r_h)=C_1 e^{-i\omega_{s,d}u}, ~~~~~~
Z^{II}_{s,d}(r<r_h)=C_1 e^{i\omega_{s,d}u},
\end{align}
where $ \omega_s $ and $ \omega_d $ represents the monochromatic frequency of the scalar and Dirac fields, respectively.

The above solutions consist a complete orthogonal family, so an arbitrary free field $ \Phi $ in the exterior region of the black hole can be expanded as
\begin{align}\label{kerrmode}
\Phi_{s,d}=\sum_{l,m}\int d\omega_{s,d}\left[a^{I}_{s,d} Z^{I}_{s,d}+a^{II}_{s,d}Z^{II}_{s,d}+H.c.  \right].
\end{align}
Similar to the Schwarzschild case, we can introduce the generalized light-like Kruskal coordinates {\color{black}\cite{L94}} $ U $ and $ V $ for the metric (\ref{ds22})
\begin{equation}
\begin{aligned}
&U=-\frac{1}{\kappa}e^{-\kappa u},~~~ V\frac{1}{\kappa}e^{\kappa v},~~~~~~ \text{for}~~~ r>r_h;\\
&U=\frac{1}{\kappa}e^{-\kappa u},~~~ V=\frac{1}{\kappa}e^{\kappa v},~~~~~~~~~ \text{for}~~~ r<r_h.
\end{aligned}
\end{equation}
Analytic expressions for all real $ U $ and $ V $ are derived by applying the Kruskal-black hole coordinate relationship.
Using Damour and Ruffini's analytic continuation method {\color{black}\cite{L95}}, their linear combination of $ Z^{I}_{s,d} $ and $ Z^{II}_{s,d} $ are analytical solutions in the whole of the spacetime
\begin{equation}
\begin{aligned}
{\Psi}^{(1)}_{s,d}\equiv&e^{\pi \omega_{s,d}/2\kappa}Z^{I}_{s,d}+e^{-\pi\omega_{s,d}/2\kappa}Z^{II*}_{s,d}\\
=&e^{\pi \omega_{s,d} r^2_h/(r_h+\alpha)}Z^{I}_{s,d}+e^{-\pi\omega_{s,d} r^2_h/(r_h+\alpha)}Z^{II*}_{s,d},
\end{aligned}
\end{equation}
\begin{equation}
\begin{aligned}
{\Psi}^{(2)}_{s,d}\equiv&e^{-\pi\omega_{s,d}/2\kappa}Z^{I*}_{s,d}+e^{\pi\omega_{s,d}/2\kappa}Z^{II}_{s,d}\\
=&e^{-\pi\omega_{s,d}r^2_h/(r_h+\alpha)}Z^{I*}_{s,d}+e^{\pi\omega_{s,d}r^2_h/(r_h+\alpha)}Z^{II}_{s,d}.
\end{aligned}
\end{equation}
In Kruskal spacetime, the quantum field $ \Phi_{s,d} $ is quantized and expanded in terms of \( {\Psi}^{(1)}_{s,d} \) and \( {\Psi}^{(2)}_{s,d} \) as the new basis
\begin{equation}
\begin{aligned}\label{smode}
\Phi_s=\sum_{l,m}\int d\omega_s \left\{2\sinh\left[2\pi\omega_s r_h^2/(r_h+\alpha)\right] \right\}^{-1/2}
\times\left[ \left(b^{(1)}_{s}\Psi^{(1)}_{s}+b^{(2)}_{s}\Psi^{(2)}_{s}\right)+H.c \right],
\end{aligned}
\end{equation}
\begin{equation}
\begin{aligned}\label{dmode}
\Phi_d=\sum_{l,m}\int d\omega_d \left\{2\cosh\left[2\pi\omega_d r_h^2/(r_h+\alpha)\right] \right\}^{-1/2}
\times \left[ \left(b^{(1)}_{d}\Psi^{(1)}_{d}+b^{(2)}_{d}\Psi^{(2)}_{d}\right)+H.c \right].
\end{aligned}
\end{equation}
Here, the quantum fields $ \Phi_s $ and $ \Phi_d $ correspond to the bosonic and fermionic fields, respectively.
Taking the inner production of states (\ref{kerrmode}) and (\ref{smode}), we can obtain the Bogoliubov transformations of bosonic field, as {\color{black}\cite{L94,L96}}
\begin{equation}
\begin{aligned}\label{RTG1}
b^{(1)}_s=\frac{a^{I}_{s}}{\sqrt{1-e^{-4\pi\omega_sr^2_h/(r_h+\alpha)}}}-\frac{a^{II\dagger}_s}{\sqrt{e^{4\pi\omega_sr^2_h/(r_h+\alpha)}-1}},\\
b^{(1)\dagger}_{s}=\frac{a^{I\dagger}_s}{\sqrt{1-e^{-4\pi\omega_sr^2_h/(r_h+\alpha)}}}-\frac{a^{II}_s}{\sqrt{e^{4\pi\omega_sr^2_h/(r_h+\alpha)}-1}},\\
b^{(2)}_{s}=\frac{a^{II}_s}{\sqrt{1-e^{-4\pi\omega_sr^2_h/(r_h+\alpha)}}}-\frac{a^{I\dagger}_s}{\sqrt{e^{4\pi\omega_sr^2_h/(r_h+\alpha)}-1}},\\
b^{(2)\dagger}_s=\frac{a^{II\dagger}_s}{\sqrt{1-e^{-4\pi\omega_sr^2_h/(r_h+\alpha)}}}-\frac{a^{I}_s}{\sqrt{e^{4\pi\omega_sr^2_h/(r_h+\alpha)}-1}},
\end{aligned}
\end{equation}
and taking the inner production of states (\ref{kerrmode})  and (\ref{dmode}), we can obtain the Bogoliubov transformations of fermionic field, as
\begin{equation}
\begin{aligned}\label{RTG2}
b^{(1)}_{d}=&\frac{a^{I}_{d}}{\sqrt{e^{-4\pi\omega_dr^2_h/(r_h+\alpha)}+1}}-\frac{a^{II\dagger}_{d}}{\sqrt{e^{4\pi\omega_dr^2_h/(r_h+\alpha)}+1}},\\
b^{(1)\dagger}=&\frac{a^{I\dagger}_{d}}{\sqrt{e^{-4\pi\omega_dr^2_h/(r_h+\alpha)}+1}}-\frac{a^{II}_{d}}{\sqrt{e^{4\pi\omega_dr^2_h/(r_h+\alpha)}+1}},\\
b^{(2)}_{d}=&\frac{a^{II}_d}{\sqrt{e^{-4\pi\omega_dr^2_h/(r_h+\alpha)}+1}}+\frac{a^{I\dagger}_{d}}{\sqrt{e^{4\pi\omega_dr^2_h/(r_h+\alpha)}+1}},    \\
b^{(2)\dagger}_{d}=&\frac{a^{II\dagger}_{d}}{\sqrt{e^{-4\pi\omega_dr^2_h/(r_h+\alpha)}+1}}+\frac{a^{I}_{d}}{\sqrt{e^{4\pi\omega_dr^2_h/(r_h+\alpha)}+1}},
\end{aligned}
\end{equation}
Here $r_h$ is a function of the dimensional parameter $\alpha$. In the limit $\alpha\to0$, $r_h\to2M$, and all coefficients reduce to their Schwarzschild counterparts. The validity of these operators requires $(r_h+\alpha)\geq0$. Accordingly, the subsequent quantum-resource calculations are restricted to $\tilde{\alpha}\geq-2$ (equivalently, $\alpha\geq-2M$), which keeps all expressions real and physically meaningful.

Using equation  (\ref{RTG1}), the Kruskal vacuum and first excited states of bosonic field in the PFDM-surrounded Schwarzschild black hole can be expressed as
\begin{eqnarray}\label{RTG3}
|0\rangle_{K}^{s}=\sqrt{1-e^{-4\pi\omega_sr^2_h/(r_h+\alpha)}}\sum^{\infty}_{n=0} e^{-2n\pi \omega_s r_h^2 / (r_h + \alpha)}|n\rangle_{I}^{s}|n\rangle_{II}^{s},
\end{eqnarray}
\begin{eqnarray}\label{RTG4}
|1\rangle_{K}^{s}=b^{(1)\dagger}_{s}|0\rangle _{K}^{s}=
\big[1-e^{-4\pi\omega_sr^2_h/(r_h+\alpha)}\big]\sum^{\infty}_{n=0}e^{-2n\pi \omega_s r_h^2 / (r_h + \alpha)}\sqrt{n+1} |n+1\rangle_{I}^{s}|n\rangle_{II}^{s},
\end{eqnarray}
where  $\{|n\rangle_{I}\}$ and $\{|n\rangle_{II}\}$ are orthonormal basis states corresponding to the exterior and interior regions of the event horizon, respectively {\color{black}\cite{L34,L52}}. Similarly, the Kruskal vacuum and first excited states of Dirac field in this spacetime can be written as
\begin{eqnarray}\label{RTG5}
|0\rangle_{K}^{d}=\frac{1}{\sqrt{e^{-4\pi\omega_dr^2_h/(r_h+\alpha)}+1}}|0\rangle_{I}^{d}|0\rangle_{II}^{d}
+\frac{1}{\sqrt{e^{4\pi\omega_dr^2_h/(r_h+\alpha)}+1}}|1\rangle_{I}^{d}|1\rangle_{II}^{d},
\end{eqnarray}
\begin{eqnarray}\label{RTG6}
|1\rangle_{K}^{d}=|1\rangle_{I}^{d}|0\rangle_{II}^{d}.
\end{eqnarray}
We emphasize that no direct nongravitational interaction between the effective-fluid source and the test quantum fields is introduced in our analysis. The effective-fluid source affects the quantum correlations through the background geometry: it changes the event-horizon radius $r_h$, the surface gravity $\kappa$, and hence the Hawking temperature $T=\kappa/(2\pi)$ and the Bogoliubov coefficients above. Hawking radiation is not treated as an additional particle species or an independently introduced environment; rather, it is encoded in the mixing between the exterior (region $I$) and interior (region $II$) modes of the same quantum field. Since an exterior observer cannot access region $II$, tracing over these modes produces a mixed reduced state and changes the entanglement and coherence accessible to Alice and Bob. This is the standard mode-access interpretation used in the Alice--Bob framework of relativistic quantum information {\color{black}\cite{L31,L32,L34,L38}}. The global state including both regions remains pure, so this reduced-state effect is distinct from the black hole information paradox. Questions concerning black hole formation, complete evaporation, backreaction, and global unitarity lie beyond the fixed-background semiclassical framework considered here.
For notational simplicity, we omit the subscripts on the frequency parameter $\omega$ in the following discussions.

\section{Quantification of quantum coherence and entanglement \label{sec3}}
A quantum state is said to exhibit coherence with respect to a chosen reference basis if it can be expressed as a nontrivial linear superposition of the basis elements. This fundamental notion of quantum coherence originates from the superposition principle, which lies at the core of quantum theory. In this work, we employ two widely used measures, the $l_1$-norm and the relative entropy of coherence, to quantify quantum coherence in the background of a Schwarzschild black hole surrounded by PFDM.
For an $n$-dimensional quantum system described by a reference basis $\{|i\rangle\}_{i=1,...,n}$, the $l_1$-norm of coherence is defined as the sum of the absolute values of the off-diagonal elements of the system's density matrix $\rho$
\begin{eqnarray}\label{wsm1}
C_{l_{1}}(\rho)=\sum_{i\neq j}|{\rho}_{i,j}|.
\end{eqnarray}
Additionally, the relative entropy of coherence (REC) quantifies the distinguishability between a given quantum state and its closest incoherent counterpart. It is defined as
\begin{eqnarray}\label{wsm2}
C_{REC}(\rho)=S(\rho_{diag})-S(\rho),
\end{eqnarray}
where $S(\rho)$ denotes the von Neumann entropy of the state $\rho$, and $\rho_{\text{diag}}$ is obtained by removing all off-diagonal elements of $\rho$, retaining only the classical probabilistic part {\color{black}\cite{L97}}.

There exist various effective methods for detecting quantum entanglement, among which the Peres-Horodecki positive partial transpose (PPT) criterion and logarithmic negativity are particularly useful due to their conceptual simplicity and computational convenience. In this work, we employ logarithmic negativity as the primary measure to quantify entanglement {\color{black}\cite{L98}}. It should be emphasized that the applicability of logarithmic negativity depends on the dimensionality of the Hilbert space. For the fermionic field considered here, the reduced density matrix corresponds to an effective two-qubit system. In this case, the PPT criterion is both necessary and sufficient for entanglement detection, and logarithmic negativity provides a complete characterization of entanglement. However, for the bosonic field, the Hilbert space is infinite-dimensional, and logarithmic negativity serves only as a sufficient but not necessary entanglement criterion.
The logarithmic negativity is defined as
\begin{eqnarray}\label{wsm3}
N(\rho)=\log_{2}\|\rho^{T}\|_{1},
\end{eqnarray}
where $\|A\|_{1}={\rm tr}\sqrt{A^{\dag}A}$ denotes the trace norm, and $\rho^{T}$
represents the partial transpose of the bipartite density matrix $\rho$ with respect to one of its parties.
Nevertheless, logarithmic negativity remains a widely adopted and computationally convenient entanglement measure. Within the present framework, it characterizes how entanglement carried by the test fields responds to the effective-fluid-modified spacetime geometry.

\section{Quantum coherence and entanglement for both fermionic and bosonic fields in the PFDM-surrounded Schwarzschild black hole \label{sec4}}
We now examine the coherence and entanglement of the fermionic and bosonic test fields in the effective-fluid-modified Schwarzschild background described above.
Following the standard observer-mode construction, Alice and Bob initially share a maximally entangled state in the asymptotically flat region. They label access to different sectors of the field modes rather than a concrete arrangement of localized detectors.
The maximally entangled state is chosen as the initial resource state because it represents the strongest quantum correlation that can be shared between two observers in the absence of gravitational effects. This choice establishes a clear reference point for quantifying how the modified black-hole geometry and Hawking radiation influence quantum resources. It should be noted that this assumption does not restrict the analysis to only idealized physical states; rather, it provides an upper bound for the preservation or degradation of quantum correlations under the gravitational background. The shared entangled state is given by
\begin{eqnarray}\label{wsm4}
|\Psi^{s/d}\rangle_{AB}=\frac{1}{\sqrt{2}}(|0\rangle_A|0\rangle_B+|1\rangle_A|1\rangle_B),
\end{eqnarray}
where the superscripts $s$ and $d$ denote the scalar (bosonic) and Dirac (fermionic) field cases, respectively. After preparing the initially entangled state, Alice remains in the asymptotically flat region, whereas Bob is restricted to the exterior modes in region $I$ near the event horizon of the effective-fluid-modified black hole.

\subsection{Bosonic coherence and entanglement }\label{sec4.1}
Using equations (\ref{RTG3}) and (\ref{RTG4}), the initially entangled state in equation (\ref{wsm4}) can be rewritten in terms of Kruskal modes for Alice and black hole modes for Bob as
\begin{align}\label{wsm5}
|\Psi^{s}\rangle_{A,I,II}&=\frac{1}{\sqrt{2}}\sum_{n=0}^{\infty}e^{-2n\pi\omega r_h^2/(r_h+\alpha)}\bigg\{\sqrt{1-e^{-4\pi\omega r_h^2/(r_h+\alpha)}}|0\rangle_{A}|n\rangle_{I}|n\rangle_{II}\notag\\
&+\sqrt{n+1}[1-e^{-4\pi\omega r_h^2/(r_h+\alpha)}]|1\rangle_{A}|n+1\rangle_{I}|n\rangle_{II}\bigg\}.
\end{align}
Accordingly, by tracing over the inaccessible modes in region $II$, we obtain the reduced density matrix for the accessible subsystem shared by Alice and Bob as
\begin{align}\label{wsm6}
\rho_{A,I}^s&=\frac{1}{2}[1-e^{-4\pi\omega r_h^2/(r_h+\alpha)}]\sum_{n=0}^{\infty}e^{-4n\pi\omega r_h^2/(r_h+\alpha)}\bigg\{|0n\rangle\langle0n|+\sqrt{(n+1)[1-e^{-4\pi\omega r_h^2/(r_h+\alpha)}]}\notag\\
&\times(|0n\rangle\langle1 n+1|+|1 n+1\rangle\langle0 n|)+(n+1)[1-e^{-4\pi\omega r_h^2/(r_h+\alpha)}]|1 n+1\rangle\langle1 n+1|\bigg\},
\end{align}
where $|nm\rangle=|n\rangle_A|m\rangle_I$.
Employing equations (\ref{wsm1}) and (\ref{wsm6}), the $l_{1}$-norm of quantum coherence of this state can be expressed as
\begin{align}\label{wsm7}
C_{l_{1}}(\rho_{A,I}^s)=[1-e^{-4\pi\omega r_h^2/(r_h+\alpha)}]^{\frac{3}{2}}\sum_{n=0}^{\infty}\sqrt{n+1}e^{-4n\pi\omega r_h^2/(r_h+\alpha)}.
\end{align}
From the equation above, we can see that thermal noise introduced by both black hole and PFDM plays a significant role in degrading coherence. Please refer to Appendix A for the calculation of the bosonic REC.

To quantify entanglement, we employ the logarithmic negativity, which involves taking the partial transpose of the density matrix $\rho_{A,I}^s$ with respect to subsystem $A$. This operation yields a new matrix, denoted as $\rho_{A,I}^{s,\rm {T}_A}$, given by
\begin{align}\label{wsm12}
\rho_{A,I}^{s,\rm {T}_A}&=\frac{1}{2}[1-e^{-4\pi\omega r_h^2/(r_h+\alpha)}]\sum_{n=0}^{\infty}e^{-4n\pi\omega r_h^2/(r_h+\alpha)}\bigg\{|0n\rangle\langle0n|+\sqrt{(n+1)[1-e^{-4\pi\omega r_h^2/(r_h+\alpha)}]}\notag\\
&\times(|1n\rangle\langle0 n+1|+|0 n+1\rangle\langle1 n|)+(n+1)[1-e^{-4\pi\omega r_h^2/(r_h+\alpha)}]|1 n+1\rangle\langle1 n+1|\bigg\},
\end{align}
which is a block matrix in the subspace of $\{|n\rangle,|n+1\rangle\}$. Therefore, the eigenvalues of $\rho_{A,I}^{s,\rm {T}_A}$ are then found to be
\begin{align}\label{wsm13}
&\frac{1}{2}[1-e^{-4\pi\omega r_h^2/(r_h+\alpha)}],\notag \\
& \frac{1}{4}[1-e^{-4\pi\omega r_h^2/(r_h+\alpha)}]e^{-4n\pi\omega r_h^2/(r_h+\alpha)}\bigg\{\mathcal{Z}\pm\sqrt{\mathcal{Z}^2+4[1-e^{-4\pi\omega r_h^2/(r_h+\alpha)}]}\bigg\},\nonumber
\end{align}
with
$$\mathcal{Z}=n[e^{4\pi \omega_s r_h^2 / (r_h + \alpha)}-1]+e^{-4\pi\omega r_h^2/(r_h+\alpha)}.$$
Therefore, the logarithmic negativity is
\begin{eqnarray}\label{wsm14}
N(\rho_{A,I}^s)=\log_2\bigg\{\frac{1}{2}[1-e^{-4\pi\omega r_h^2/(r_h+\alpha)}]+\frac{1}{2}[1-e^{-4\pi\omega r_h^2/(r_h+\alpha)}]\sum_{n=0}^{\infty}e^{-4n\pi\omega r_h^2/(r_h+\alpha)}
\times\sqrt{\mathcal{Z}^2+4[1-e^{-4\pi\omega r_h^2/(r_h+\alpha)}]}\bigg\}.
\end{eqnarray}

\begin{figure}[htbp]
\centering
\includegraphics[height=1.8in,width=2.0in]{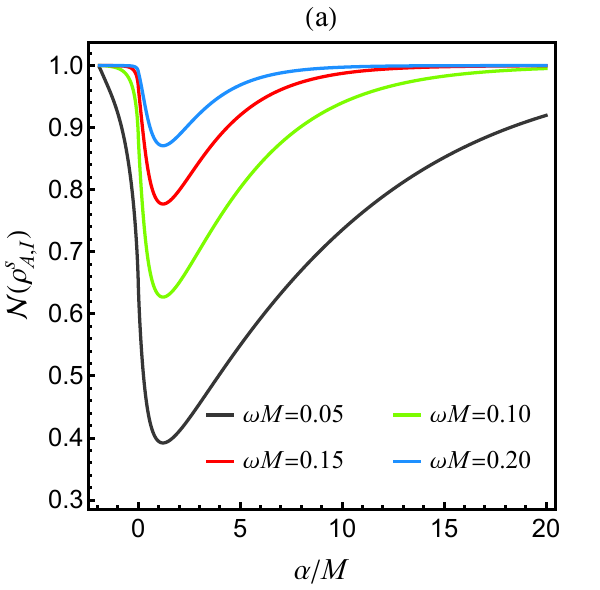}
\includegraphics[height=1.8in,width=2.0in]{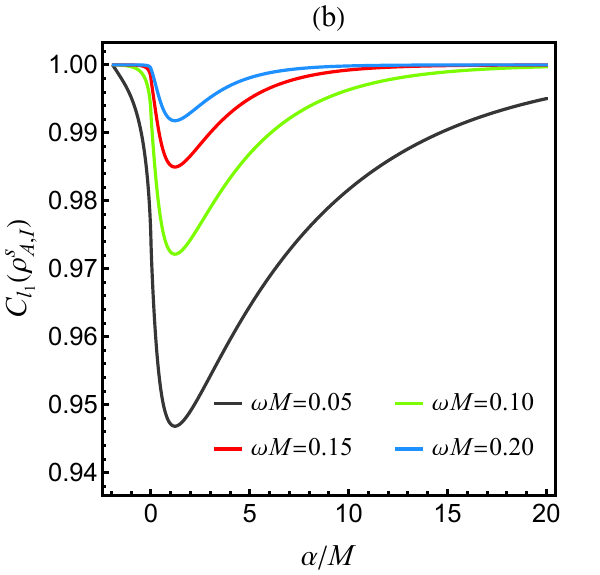}
\includegraphics[height=1.8in,width=2.0in]{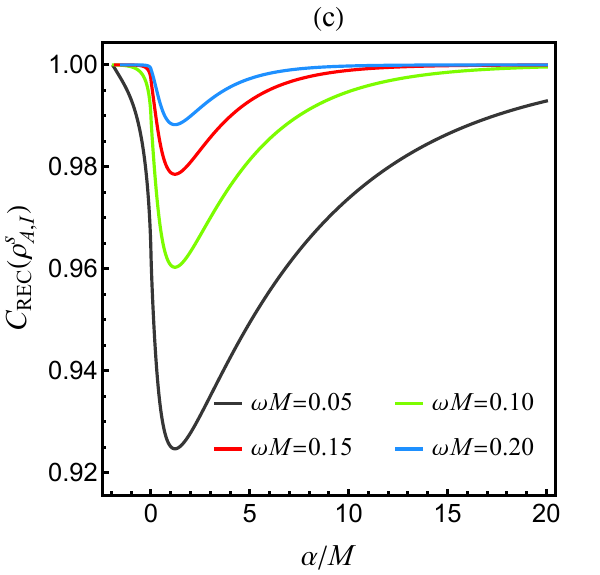}
\caption{Bosonic entanglement and coherence as functions of the dimensionless parameter $\tilde{\alpha}$. The black, green, red, and blue curves correspond to $\omega M=0.05,0.1,0.15,0.2$, respectively. The interval $-2\leq\tilde{\alpha}<0$ is the positive-density PFDM sector, whereas $\tilde{\alpha}>0$ is shown as a formal extension of the metric family.}\label{Fig3}
\end{figure}

In Fig.\ref{Fig3}, we present a detailed analysis of bosonic entanglement and coherence as functions of the dimensionless parameter $\tilde{\alpha}$ for various values of $\omega M$. For a given $\tilde{\alpha}$, both the entanglement and coherence of the state $\rho_{A,I}^s$ are lower when $\omega M$ is smaller. This is because, for fixed frequency $\omega$, a smaller $\omega M$ corresponds to a smaller  mass of the black hole $M$. A lower $M$ results in a higher Hawking temperature and an increased rate of particle creation via the Hawking effect, which in turn leads to a more significant degradation of bosonic entanglement and coherence. Therefore, the degradation of quantum entanglement and coherence is more pronounced in the presence of lighter black holes. Furthermore, for fixed $M$, both bosonic entanglement and coherence increase monotonically with $\omega$, indicating that higher-frequency field modes are more robust against decoherence and more effective in preserving quantum entanglement and coherence  near the event horizon.

Fig.\ref{Fig3} also makes the two parameter sectors transparent. Within the positive-density PFDM sector $-2\leq\tilde{\alpha}<0$, bosonic entanglement and coherence decrease monotonically as $\tilde{\alpha}$ increases. The recovery occurs only after the parameter is continued into the formal branch $\tilde{\alpha}>0$; consequently, the nonmonotonic behavior belongs to the complete extended metric family and is not interpreted as a conventional dark-matter effect. Geometrically, this behavior follows from the variation of the Hawking temperature: as $\tilde{\alpha}$ increases over the complete range, the temperature first rises and then falls. The initial rise strengthens particle creation and thermal decoherence, whereas the subsequent fall suppresses particle creation and allows the quantum resources to recover. For larger black holes, the dependence within $-2\leq\tilde{\alpha}\leq0$ becomes extremely weak because the Hawking temperature, and hence the effective thermal noise experienced by the test fields, is small. The PFDM-induced geometric response of the quantum resources is therefore most pronounced for smaller black holes.

\subsection{Fermionic coherence and entanglement  }\label{sec4.2}
Similar to the treatment of the maximally entangled bosonic state, we rewrite the fermionic entangled state in equation (\ref{wsm4}) using the black hole modes defined in equations  (\ref{RTG5}) and (\ref{RTG6}) as
\begin{eqnarray}\label{wsm15}
|\Psi^{d}\rangle_{A,I,II}=\frac{1}{\sqrt{2}}\bigg\{\frac{1}{\sqrt{e^{-4\pi \omega_s r_h^2 / (r_h + \alpha)}+1}}|0\rangle_A|0\rangle_{I}|0\rangle_{II}+\frac{1}{\sqrt{e^{4\pi \omega_s r_h^2 / (r_h + \alpha)}+1}}|0\rangle_A|1\rangle_{I}|1\rangle_{II}
+|1\rangle_A|1\rangle_{I}|0\rangle_{II}\bigg\} .
\end{eqnarray}
Since Bob cannot access the region inside the event horizon, we trace over the inaccessible modes in region $II$ to obtain the reduced density matrix in the exterior region
\begin{eqnarray}\label{wsm16}
\nonumber \rho_{A,I}^d&=&\frac{1}{2}\bigg\{\frac{1}{e^{-4\pi \omega r_h^2 / (r_h + \alpha)}+1}|00\rangle\langle 00|+\frac{1}{e^{4\pi \omega r_h^2 / (r_h + \alpha)}+1}|01\rangle\langle 01|\\
&+&\frac{1}{\sqrt{e^{-4\pi \omega r_h^2 / (r_h + \alpha)}+1}}|11\rangle\langle 00|+\frac{1}{\sqrt{e^{-4\pi \omega r_h^2 / (r_h + \alpha)}+1}}|00\rangle\langle 11|+|11\rangle\langle 11|\bigg\}.
\end{eqnarray}
Based on the definition of quantum coherence, the $l_{1}$-norm of fermionic coherence is calculated as
\begin{eqnarray}\label{wsm17}
C_{l_{1}}(\rho_{A,I}^d)=\frac{1}{\sqrt{e^{-4\pi \omega r_h^2 / (r_h + \alpha)}+1}}.
\end{eqnarray}
Please refer to Appendix A for the calculation of the fermionic REC.

We now compute the logarithmic negativity of the state $\rho_{A,I}^d$. Performing the partial transpose of $\rho_{A,I}^d$ with respect to subsystem $A$, we obtain
\[
\rho_{A,I}^{d,\rm {T}_A}=\frac{1}{2}\begin{pmatrix}
\frac{1}{e^{-4\pi \omega r_h^2 / (r_h + \alpha)}+1} & 0 & 0 & 0 \\
0 & \frac{1}{e^{4\pi \omega r_h^2 / (r_h + \alpha)}+1} & \frac{1}{\sqrt{e^{-4\pi \omega r_h^2 / (r_h + \alpha)}+1}} & 0 \\
0 & \frac{1}{\sqrt{e^{-4\pi \omega r_h^2 / (r_h + \alpha)}+1}} & 0 & 0 \\
0 & 0 & 0 & 1
\end{pmatrix}.
\]
The eigenvalues of $\rho_{A,I}^{d,\rm {T}_A}$ are $(1,1,\frac{1}{e^{-4\pi \omega r_h^2 / (r_h + \alpha)}+1},\frac{-1}{e^{-4\pi \omega r_h^2 / (r_h + \alpha)}+1})/2$, where the negative eigenvalue signals the presence of entanglement, as negativity in the partial transpose is a well-established entanglement witness.
Using equation (\ref{wsm3}), the fermionic logarithmic negativity is therefore given by
\begin{eqnarray}\label{wsm18}
 \nonumber N(\rho_{A,I}^d)=\log_2 \bigg[1+\frac{1}{e^{-4\pi \omega r_h^2 / (r_h + \alpha)}+1}\bigg].
\end{eqnarray}

\begin{figure}[htbp]
\centering
\includegraphics[height=1.8in,width=2.0in]{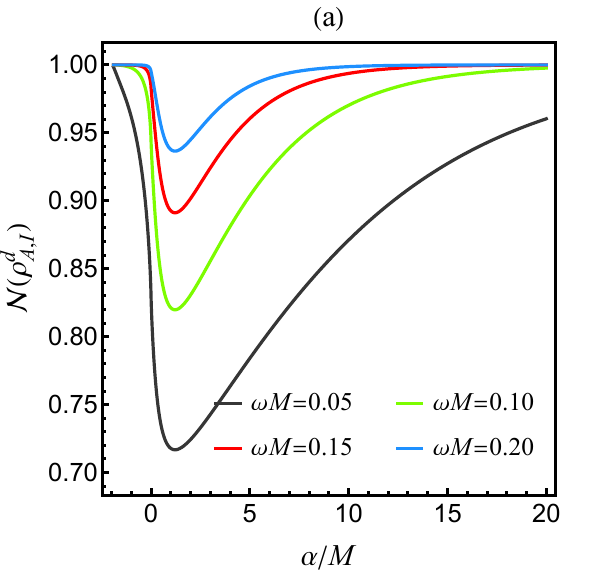}
\includegraphics[height=1.8in,width=2.0in]{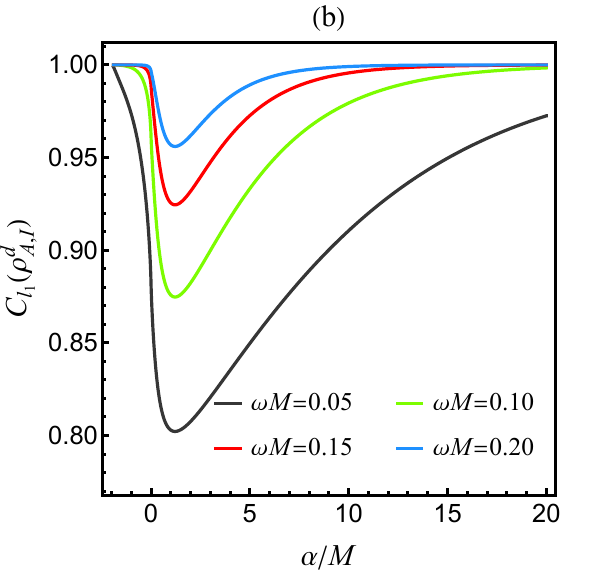}
\includegraphics[height=1.8in,width=2.0in]{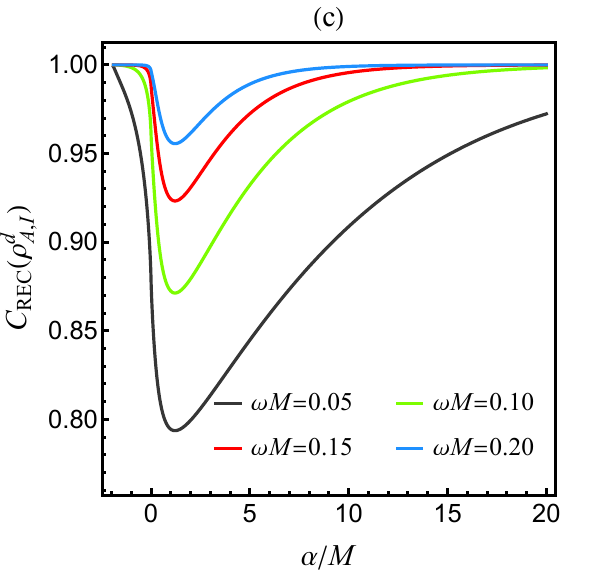}
\caption{Fermionic entanglement and coherence as functions of the dimensionless parameter $\tilde{\alpha}$. The black, green, red, and blue curves correspond to $\omega M=0.05$, $0.1$, $0.15$, and $0.2$, respectively. The interval $-2\leq\tilde{\alpha}<0$ is the positive-density PFDM sector, whereas $\tilde{\alpha}>0$ is the formal extension of the metric family.}\label{Fig4}
\end{figure}

Fig.\ref{Fig4} illustrates the fermionic entanglement and coherence as functions of $\tilde{\alpha}$ for different values of $\omega M$. Their dependence on the two parameter sectors parallels that of the bosonic quantities discussed above: the positive-density PFDM sector is monotonic, while the recovery occurs in the formally extended branch. Across the complete parameter range, fermionic entanglement consistently exceeds bosonic entanglement, showing that bosonic entanglement is more susceptible to the Hawking-induced geometric effect. Conversely, fermionic coherence remains lower than bosonic coherence and therefore exhibits the stronger relative response. These comparisons do not constitute a direct dark-matter detection proposal; rather, they identify which combinations of field statistics and quantum resource are most sensitive to changes in the effective-fluid-modified spacetime geometry.

\section{Conclusions\label{GSCDGE}}
In this work, we have investigated the entanglement and coherence of massless scalar and Dirac test fields in the fixed background of a Schwarzschild black hole surrounded by the effective source conventionally termed PFDM. These fields encode quantum information and are not assumed to be microscopic dark-matter constituents. The effective source acts only through its modification of the event horizon, surface gravity, Hawking temperature, and the resulting Bogoliubov mode mixing.

Our results must be interpreted separately in the two parameter sectors. Within the positive-density PFDM sector $-2\leq\tilde{\alpha}<0$, entanglement and coherence decrease monotonically as $\tilde{\alpha}$ increases, although this dependence becomes extremely weak for sufficiently large black holes. The recovery of these quantum resources occurs only in the formal extension $\tilde{\alpha}>0$. Thus, the nonmonotonic behavior over the complete parameter range is a property of the extended metric family and is not assigned a conventional dark-matter interpretation. The underlying mechanism is the corresponding nonmonotonic variation of the Hawking temperature, which controls the thermal mode mixing experienced by the test fields.

Bosonic entanglement is more susceptible to the geometric effect than fermionic entanglement, whereas fermionic coherence displays a stronger relative response than bosonic coherence. Here, ``probing'' describes the theoretical sensitivity of quantum resources to the assumed geometry. Observational studies of dark-matter pressure and density provide guidance for assessing astrophysical applications, as discussed in Section II~{\color{black}\cite{Potapov2016,Kumar2019}}. Once the background parameters are constrained within an appropriate description of the matter profile and stresses, these constraints would restrict the predicted entanglement and coherence through the Hawking temperature $T=(r_h+\alpha)/(4\pi r_h^2)$ and the ratio $\omega/T$. When the resulting temperature is close to the Schwarzschild value, the quantum measures likewise approach their Schwarzschild results at fixed $M$ and $\omega$. Extending the analysis to more realistic halo profiles or alternative effective geometries~{\color{black}\cite{L73,L74,ADO4}} would help assess the robustness of these predictions. Connecting them to measurable signals would additionally require a localized detector model with specified trajectories, field--detector couplings, and a measurement protocol. The present results thus describe the response of test fields to a phenomenological background, with astrophysical applications requiring observational validation of the background.

\begin{acknowledgments}
S. M. Wu acknowledges support from the National Natural Science Foundation of China (Grant Nos. 12575056 and 12205133), the Natural Science Foundation of the Liaoning Scientific Committee (Grant
No. 2026-MS-281), and  LiaoNing Revitalization Talents Program (XLYC2503099).
W. Liu acknowledges support from the China Postdoctoral Science Foundation (Grant No. 2025M783393) and the National Natural Science Foundation of China (Grant Nos. 12547147 and 12605108).
\end{acknowledgments}

\appendix

\section{REC in the PFDM-surrounded Schwarzschild black hole}
\label{app:steering}

To compute the REC for the state $\rho_{A,I}^s$, we first determine its eigenvalues. The density matrix $\rho_{A,I}^s$ possesses a block-diagonal structure composed of $2 \times 2$ submatrices $\Delta_n$ along the diagonal, with all off-diagonal elements vanishing. It can be expressed as
\begin{eqnarray}\label{wsm8}
\rho_{A,I}^s=\frac{1}{2}[1-e^{-4\pi\omega r_h^2/(r_h+\alpha)}]
\begin{pmatrix}
0 &  &  &  & \\
 & \Delta_0 &  &  & \\
 &  & \Delta_1 &  & \\
 &  &  & \ddots & \\
 &  &  &  & \Delta_n \\
 &  &  &  &  & \ddots
\end{pmatrix},
\end{eqnarray}
where each block $\Delta_n$ takes the form
\begin{eqnarray}\label{wsm9}
\Delta_n(\rho_{A,I}^s)=e^{-4n\pi\omega r_h^2/(r_h+\alpha)}
\begin{pmatrix}
1 & \sqrt{(n+1)[1-e^{-4\pi\omega r_h^2/(r_h+\alpha)}]} \\
\sqrt{(n+1)[1-e^{-4\pi\omega r_h^2/(r_h+\alpha)}]} & (n+1)[1-e^{-4\pi\omega r_h^2/(r_h+\alpha)}] \nonumber
\end{pmatrix},
\end{eqnarray}
highlighting the matrix's sparse and well-organized structure, which facilitates both analytical treatment and numerical implementation.
The eigenvalues of this block $n$-th in the state $\rho_{A,I}^s$ are given by
\begin{align}\label{wsm10}
\Lambda_n=\frac{1}{2}[1-e^{-4\pi\omega r_h^2/(r_h+\alpha)}]e^{-4n\pi\omega r_h^2/(r_h+\alpha)}\bigg\{1+(n+1)[1-e^{-4\pi\omega r_h^2/(r_h+\alpha)}] \bigg\}.
\end{align}
The total trace of the density matrix over all blocks satisfies the normalization condition $\mathrm{Tr}(\rho_{A,I}^s)=\sum_{n = 0}^{\infty}\Lambda_n=1$. Using equations (\ref{wsm2}) and (\ref{wsm10}), the analytical expression for the REC is obtained as
\begin{align}\label{wsm11}
C_{REC}(\rho_{A,I}^s)&=\sum_{n=0}^{\infty}\bigg\{\Lambda_n\log_2\Lambda_n-
\mathcal{P}_n\log_2\mathcal{P}_n-\mathcal{Q}_n\log_2\mathcal{Q}_n\bigg\},
\end{align}
with $$\mathcal{P}_n=\frac{1}{2}[1-e^{-4\pi\omega r_h^2/(r_h+\alpha)}]e^{-4n\pi\omega r_h^2/(r_h+\alpha)},$$
and
 $$\mathcal{Q}_n=\frac{1}{2}[1-e^{-4\pi\omega r_h^2/(r_h+\alpha)}]^2e^{-4n\pi\omega r_h^2/(r_h+\alpha)}(n+1).$$

The density operator of equation (\ref{wsm16}) can be written in matrix form as
\[
\rho_{A,I}^d=\frac{1}{2}\begin{pmatrix}
\frac{1}{e^{-4\pi \omega r_h^2 / (r_h + \alpha)}+1} & 0 & 0 & \frac{1}{\sqrt{e^{-4\pi \omega r_h^2 / (r_h + \alpha)}+1}} \\
0 & \frac{1}{e^{4\pi \omega r_h^2 / (r_h + \alpha)}+1} & 0 & 0 \\
0 & 0 & 0 & 0 \\
\frac{1}{\sqrt{e^{-4\pi \omega r_h^2 / (r_h + \alpha)}+1}} & 0 & 0 & 1
\end{pmatrix},
\]
in the basis $\{|00\rangle,|01\rangle,|10\rangle,|11\rangle\}$,
with non-zero eigenvalues given by
\[
\frac{e^{-4\pi\omega r_h^2/(r_h+\alpha)}}{2+2e^{-4\pi\omega r_h^2/(r_h+\alpha)}},\quad \frac{2+e^{-4\pi\omega r_h^2/(r_h+\alpha)}}{2+2e^{-4\pi\omega r_h^2/(r_h+\alpha)}}.
\]
The REC is given by
\begin{eqnarray}\label{wsm18}
 \nonumber C_{REC}(\rho_{A,I}^d)&=&\frac{e^{-4\pi\omega r_h^2/(r_h+\alpha)}}{2+2e^{-4\pi\omega r_h^2/(r_h+\alpha)}}\log_2\bigg[\frac{e^{-4\pi\omega r_h^2/(r_h+\alpha)}}{2+2e^{-4\pi\omega r_h^2/(r_h+\alpha)}}\bigg]+\frac{2+e^{-4\pi\omega r_h^2/(r_h+\alpha)}}{2+2e^{-4\pi\omega r_h^2/(r_h+\alpha)}}\\ \nonumber
&\times&\log_2\bigg[\frac{2+e^{-4\pi\omega r_h^2/(r_h+\alpha)}}{2+2e^{-4\pi\omega r_h^2/(r_h+\alpha)}}\bigg]-\frac{1}{2[e^{-4\pi \omega r_h^2 / (r_h + \alpha)}+1]}\log_2\bigg\{\frac{1}{2[e^{-4\pi \omega r_h^2 / (r_h + \alpha)}+1]}\bigg\}\\  \nonumber
&-&\frac{1}{2[e^{4\pi \omega r_h^2 / (r_h + \alpha)}+1]}\log_2\bigg\{\frac{1}{2[e^{4\pi \omega r_h^2 / (r_h + \alpha)}+1]}\bigg\}+\frac{1}{2}.
\end{eqnarray}


\begin{thebibliography}{999}
\color{black}\hypersetup{urlcolor=black,citecolor=black}
\bibitem{L1}
J. Akin, Y. Zhao, P. G. Kwiat, E. A. Goldschmidt, and K. Fang, Faithful Quantum Teleportation via a Nanophotonic Nonlinear Bell State Analyzer, Phys. Rev. Lett. {\bf134}, 160802 (2025).

\bibitem{L2}
J. Zhao, \textit{et al.}, Enhancing quantum teleportation efficacy with noiseless linear amplification,  Nat. Commun. {\bf14}, 4745  (2023).

\bibitem{L3}
J. Grebel, \textit{et al.}, Bidirectional Multiphoton Communication between Remote Superconducting Nodes, Phys. Rev. Lett. {\bf132}, 047001 (2024).

\bibitem{L4}
A. Z. Ding, \textit{et al.}, Quantum Control of an Oscillator with a Kerr-cat Qubit, Nat. Commun. {\bf16}, 5279 (2025).

\bibitem{L5}
C. Zhang, \textit{et al.}, Experimental Side-Channel-Secure Quantum Key Distribution, Phys. Rev. Lett. {\bf128}, 190503 (2022).

\bibitem{L6}
Y. Li, \textit{et al.}, Microsatellite-based real-time quantum key distribution, Nature {\bf640}, 47  (2025).

\bibitem{L7}
N. Gisin, G. Ribordy, W. Tittel, and H. Zbinden, Quantum cryptography, Rev. Mod. Phys. {\bf74}, 145 (2002).

\bibitem{L8}
Y. Zhao, \textit{et al.}, Direct photo-patterning of halide perovskites toward machine-learning-assisted erasable photonic cryptography,  Nat. Commun.  {\bf16}, 3316 (2025).

\bibitem{L9}
R. Horodecki, P. Horodecki, M. Horodecki, and K. Horodecki, Quantum entanglement, Rev. Mod. Phys. {\bf81}, 865 (2009).

\bibitem{L10}
E. Chitambar, G. Gour, Quantum resource theories, Rev. Mod. Phys. {\bf91}, 025001 (2019).

\bibitem{L11}
A. J. Leggett, Macroscopic quantum systems and the quantum theory of measurement, Prog. Theor. Phys. Suppl. {\bf69}, 80 (1980).

\bibitem{L12}
F. Bibak, F. D. Santo, and B. Daki\'{c}, Quantum Coherence in Networks, Phys. Rev. Lett. {\bf133}, 230201 (2024).

\bibitem{L13}
H. L. Shi, S. Ding, Q. K. Wan, X. H. Wang, and W. L. Yang, Entanglement, Coherence, and Extractable Work in Quantum Batteries, Phys. Rev. Lett. {\bf129}, 130602 (2022).

\bibitem{L14}
F. Ahnefeld, T. Theurer, D. Egloff, J. M. Matera, and M. B. Plenio, Coherence as a Resource for Shor's Algorithm, Phys. Rev. Lett. {\bf129}, 120501 (2022).

\bibitem{L15}
Y. Karli, \textit{et al.}, Controlling the photon number coherence of solid-state quantum light sources for quantum cryptography, Npj Quantum Inf. {\bf10},  17 (2024).

\bibitem{L16}
A. Yamauchi, S. Fujiwara, N. Kimizuka, \textit{et al.}, Modulation of triplet quantum coherence by guest-induced structural changes in a flexible metal-organic framework. Nat. Commun. {\bf15}, 7622 (2024).

\bibitem{L17}
Y. Wang, Y. Hu, J. P. Guo, J. Gao, B. Song, L. Jiang, A physical derivation of high-flux ion transport in biological channel via quantum ion coherence,  Nat. Commun. {\bf15},  7189 (2024).

\bibitem{L18}
A. Streltsov, G. Adesso, and M. B. Plenio, \emph{Colloquium}: Quantum coherence as a resource, Rev. Mod. Phys. {\bf89}, 041003 (2017).

\bibitem{L19}
C. Cepollaro, \textit{et al.}, Sum of Entanglement and Subsystem Coherence Is Invariant under Quantum Reference Frame Transformations,  Phys. Rev. Lett. {\bf135}, 010201 (2025).

\bibitem{L20}
H. J. Kim, S. Lee, Relation between quantum coherence and quantum entanglement in quantum measurements, Phys. Rev. A {\bf106}, 022401 (2022).

\bibitem{L21}
J. Borregaard, I. Pikovski, Testing quantum theory on curved spacetime with quantum networks, Phys. Rev. Res. {\bf7}, 023192 (2025).

\bibitem{L22}
H. Du, R. B. Mann, Fisher information as a probe of spacetime structure: relativistic quantum metrology in (A)dS. J. High Energy Phys. {\bf2021}, 112 (2021).

\bibitem{L23}
A. Roura, Quantum probe of space-time curvature, Science {\bf375}, 142--143 (2022).

\bibitem{L24}
Y. Yang, J. Jing, Z. Tian, Probing cosmic string spacetime through parameter estimation. Eur. Phys. J. C {\bf82}, 688 (2022).

\bibitem{L33}
T. Gonzalez-Raya, S. Pirandola and M. Sanz, Satellite-based entanglement distribution and quantum teleportation with continuous variables, Commun. Phys. {\bf7},  126 (2024).

\bibitem{L25}
Y. Ma, \textit{et al.}, Proposal for gravitational-wave detection beyond the standard quantum limit through EPR entanglement, Nat. Phys. {\bf13}, 776 (2017).

\bibitem{L26}
M. Tse, \textit{et al.}, Quantum-Enhanced Advanced LIGO Detectors in the Era of Gravitational-Wave Astronomy, Phys. Rev. Lett. {\bf123}, 231107 (2019).

\bibitem{L27}
S. M. Wu, R. D. Wang, X. L. Huang, Z. Wang, Does gravitational wave assist vacuum steering and Bell nonlocality?, J. High Energy Phys.  {\bf2024},  155 (2024).

\bibitem{L28}
M. Parikh, F. Wilczek, and G. Zahariade, Quantum Mechanics of Gravitational Waves, Phys. Rev. Lett. {\bf127}, 081602 (2021).

\bibitem{L29}
S. Barman, I. Chakraborty, S. Mukherjee, Signatures of gravitational wave memory in the radiative process of entangled quantum probes, Phys. Rev. D {\bf111}, 025021 (2025).

\bibitem{L30}
 S. Barman, I. Chakraborty, S. Mukherjee, Entanglement harvesting for different gravitational wave burst profiles with and without memory, J. High Energy Phys. {\bf2023}, 180 (2023).

\bibitem{L31}
I. Fuentes-Schuller and R. B. Mann, Alice falls into a black hole: Entanglement in noninertial frames, Phys. Rev. Lett. {\bf95}, 120404 (2005).

\bibitem{L32}
P. M. Alsing, I. Fuentes-Schuller, R. B. Mann and T. E. Tessier, Entanglement of Dirac fields in noninertial frames, Phys. Rev. A {\bf74}, 032326 (2006).

\bibitem{L40}
Q. Liu, T. Liu, C. Wen, and J. Wang, Optimal quantum strategy for locating Unruh channels, Phys. Rev. A {\bf110}, 022428 (2024).

\bibitem{L43}
S. Banerjee, A. K. Alok, S. Omkar and R. Srikanth, Characterization of Unruh channel in the context of open quantum systems, J. High Energy Phys. {\bf2017}, 82 (2017).

\bibitem{L45}
H. M. Reji, H. S. Hegde and R. Prabhu, Conditions for separability in multiqubit systems with an accelerating qubit using a conditional entropy, Phys. Rev. A {\bf110}, 032403 (2024).

\bibitem{L50}
S. Harikrishnan, S. Jambulingam, P. P. Rohde and C. Radhakrishnan, Accessible and inaccessible quantum coherence in relativistic quantum systems, Phys. Rev. A {\bf105}, 052403 (2022).

\bibitem{L52}
D. E. Bruschi, J. Louko, E. Mart\'{\i}n-Mart\'{\i}nez, A. Dragan, I. Fuentes, Unruh effect in quantum information beyond the single-mode approximation, Phys. Rev. A {\bf82}, 042332 (2010).

\bibitem{L53}
D. E. Bruschi, A. Dragan, I. Fuentes, J. Louko, Particle and antiparticle bosonic entanglement in noninertial frames,
Phys. Rev. D {\bf86}, 025026 (2012).

\bibitem{LL70}
J. K. Basak, D. Giataganas, S. Mondal and W. Y. Wen, Reflected entropy and Markov gap in noninertial frames, Phys. Rev. D {\bf108}, 125009 (2023).

\bibitem{L34}
Q. Pan and J. Jing, Hawking radiation, entanglement, and teleportation in the background of an asymptotically flat static black hole, Phys. Rev. D {\bf78}, 065015 (2008).

\bibitem{L35}
S. M. Wu, X. W. Fan, R. D. Wang, H. Y. Wu, X. L. Huang and H. S. Zeng, Does Hawking effect always degrade fidelity of quantum teleportation in Schwarzschild spacetime?, J. High Energy Phys. {\bf11}, 232 (2023).

\bibitem{L36}
A. Ali, S. Al-Kuwari, M. Ghominejad, M. T. Rahim, D. Wang and S. Haddadi, Quantum characteristics near event horizons, Phys. Rev. D {\bf110}, 064001 (2024).

\bibitem{L38}
E. Mart\'{\i}n-Mart\'{\i}nez, L. J. Garay and J. Le\'{o}n, Unveiling quantum entanglement degradation near a Schwarzschild black hole,  Phys. Rev. D {\bf82}, 064006 (2010).

\bibitem{L47}
S. M. Wu, X. W. Teng, J. X. Li, S. H. Li, T. H. Liu and J. C. Wang, Genuinely accessible and inaccessible entanglement in Schwarzschild black hole, Phys. Lett. B {\bf848}, 138334 (2024).

\bibitem{L48}
T. Zhang, X. Wang and S. M. Fei, Hawking effect can generate physically inaccessible genuine tripartite nonlocality, Eur. Phys. J. C {\bf83}, 607 (2023).

\bibitem{L49}
S. M. Wu and H. S. Zeng, Genuine tripartite nonlocality and entanglement in curved spacetime, Eur. Phys. J. C {\bf82}, 4 (2022).

\bibitem{Wu:2025euf}
S. M. Wu, X. W. Teng, W. M. Li, Y. X. Wang and J. Lu, Nonseparability of multipartite systems in dilaton black~hole, JCAP \textbf{09}, 030  (2025).

\bibitem{L37}
C. Y. Liu, Z. W. Long and Q. L. He, Quantum coherence and quantum Fisher information of Dirac particles in curved spacetime under decoherence, Phys. Lett. B {\bf857}, 138991 (2024).

\bibitem{L39}
S. Elghaayda, X. Zhou, M. Mansour, Distribution of distance-based quantum resources outside a radiating Schwarzschild black hole, Class. Quantum Grav. {\bf41}, 195010 (2024).

\bibitem{L51}
W. M. Li, S. M. Wu, Bosonic and fermionic coherence of N-partite states in the background of a dilaton black hole, J. High Energy Phys. {\bf2024},  144 (2024).

\bibitem{L60}
H. Dolatkhah, A. Czerwinski, A. Ali, S. Al-Kuwari and S. Haddadi, Tripartite measurement uncertainty in Schwarzschild space-time, Eur. Phys. J. C \textbf{84}, 1162 (2024).

\bibitem{L61}
S. Haddadi, M. A. Yurischev, M. Y. Abd-Rabbou, M. Azizi, M. R. Pourkarimi and M. Ghominejad, Quantumness near a Schwarzschild black hole, Eur. Phys. J. C \textbf{84}, 42 (2024).

\bibitem{Li:2025jlu}
W. M. Li, J. Lu and S. M. Wu, Multiqubit coherence of mixed states near event horizon, J. Cosmol. Astropart. Phys. {\bf02}, 058 (2026).

\bibitem{L62}
R. Li, Z. Zhao, Entanglement harvesting of circularly accelerated detectors with a reflecting boundary,  J. High Energy Phys. {\bf2025}, 185 (2025).

\bibitem{L63}
S. H.  Li, S. H. Shang, S. M. Wu, Does acceleration always degrade quantum entanglement for tetrapartite Unruh-DeWitt detectors?, J. High Energy Phys. {\bf2025}, 214 (2025).

\bibitem{L69}
Z. Liu, J. Zhang, R. B. Mann, and H. Yu, Does acceleration assist entanglement harvesting?, Phys. Rev. D {\bf105}, 085012  (2022).

\bibitem{Liu:2015oat}
X. Liu, Z. Tian, J. Wang and J. Jing, Protecting quantum coherence of two-level atoms from vacuum fluctuations of electromagnetic field, Annals Phys. \textbf{366}, 102-112 (2016).

\bibitem{Liu:2025bzv}
X. Liu, Z. Tian and J. Jing, Dissipation suppression for an Unruh-DeWitt battery with a reflecting boundary, Sci. China Phys. Mech. Astron. \textbf{68}, no.10, 100412 (2025).

\bibitem{L46}
S. Elghaayda, A. Ali, M. Y. Abd-Rabbou, M. Mansour, S. Al-Kuwari,  Quantum correlations and metrological advantage among Unruh-DeWitt detectors in de Sitter spacetime, Eur. Phys. J. C {\bf85}, 447 (2025).

\bibitem{L54}
S. M. Wu, C. X. Wang, D. D. Liu, X. L. Huang, H. S. Zeng, Would quantum coherence be increased by curvature effect in de Sitter space?, J. High Energy Phys. {\bf2023}, 115  (2023).

\bibitem{L56}
Z. D. Wei, W. Han, Y. J. Zhang, Z. X. Man, Y. J. Xia, H. Fan, Effect of the gravitational redshift on the precision of phase estimation, Phys. Rev. D \textbf{111},  026007 (2025).

\bibitem{L57}
Z. Tian, X. Liu, J. Wang, J. Jing, Dissipative dynamics of an open quantum battery in the BTZ spacetime,  J. High Energy Phys. {\bf2025}, 188 (2025).

\bibitem{L58}
X. Liu, W. Liu, Z. Liu, J. Wang, Harvesting correlations from BTZ black hole coupled to a Lorentz-violating vector field, J. High Energy Phys. {\bf08}, 094 (2025).

\bibitem{L64}
W. Izquierdo,  J. Beltran,  E. Arias, Enhancement of harvesting vacuum entanglement in Cosmic String Spacetime, J. High Energy Phys. {\bf2025}, 49 (2025).

\bibitem{L65}
T. Rick Perche, J. Polo-G\'{o}mez, B. de S. L. Torres,  E. Mart\'{\i}n-Mart\'{\i}nez, Fully relativistic entanglement harvesting,  Phys. Rev. D {\bf109}, 045018 (2024).

\bibitem{L66}
Y. Ji, J. Zhang, H. Yu, Entanglement harvesting in cosmic string spacetime, J. High Energy Phys. {\bf2024}, 161 (2024).

\bibitem{L59}
X. Liu, C. Zeng and J. Wang, Generation of quantum entanglement in superposed diamond spacetime, Eur. Phys. J. C  {\bf85}, 539 (2025).

\bibitem{L70}
A. Chakraborty,  L. Hackl,  M. Zych, Entanglement harvesting in quantum superposed spacetime, Phys. Rev. D {\bf111}, 104052 (2025).

\bibitem{Tang:2025mtc}
Y. Tang, W. Liu, Z. Liu, and J. Wang,
Can the signatures of quantum superposition be detected through correlation harvesting?,
J. High Energy Phys. {\bf01}, 045 (2026).

\bibitem{L41}
Y. Tang, W. Liu, J. Wang, Observational signature of Lorentz violation in acceleration radiation, Eur. Phys. J. C {\bf85}, 1108 (2025).

\bibitem{L42}
S. Sen, A. Mukherjee and S. Gangopadhyay, Entanglement degradation as a tool to detect signatures of modified gravity, Phys. Rev. D {\bf109}, 046012 (2024).

\bibitem{L44}
X. Liu, Z. Tian, J. Jing, Entanglement dynamics in $\kappa$-deformed spacetime, Sci. China Phys. Mech. Astron. {\bf67}, 100411 (2024).

\bibitem{L55}
W. Liu, C. Wen, J. Wang, Lorentz violation alleviates gravitationally induced entanglement degradation, J. High Energy Phys. {\bf2025}, 184 (2025).

\bibitem{AraujoFilho:2025nmc}
A. A. Ara\'ujo Filho and W. Liu,
Entanglement, equivalence principle, and HBAR entropy in a new bumblebee black hole,
arXiv:2512.17567 [gr-qc].

\bibitem{Liu:2025hcx}
X. Liu, W. Liu, and S. M. Wu,
Entanglement degradation of static black holes in effective quantum gravity,
Phys. Lett. B {\bf875}, 140334 (2026).

\bibitem{L67}
Z. Liu, R. Q. Yang, H. Fan, J. Wang, Simulation of the massless Dirac field in 1+1D curved spacetime, Sci. China Phys. Mech. Astron. {\bf68}, 290411 (2025).

\bibitem{L68}
W. Liu, D. Wu, J. Wang, Light rings and shadows of static black holes in effective quantum gravity, Phys. Lett. B {\bf858}, 139052  (2024).

\bibitem{Liu:2025zts}
Z. Liu, W. Liu, X. Liu, and J. Wang,
Wormhole-induced correlation: A link between two universes,
Phys. Lett. B {\bf879}, 140667 (2026).

\bibitem{L71}
G. Bertone, D. Hooper, History of dark matter, Rev. Mod. Phys. {\bf90}, 045002 (2018).

\bibitem{L72}
A. Arbey, F. Mahmoudi, Dark matter and the early Universe: A review, Prog. Part. Nucl. Phys. {\bf119}, 103865 (2021).

\bibitem{L73}
J. F. Navarro, C. S. Frenk, and S. D. M. White, The Structure of Cold Dark Matter Halos, Astrophys. J. {\bf462}, 563 (1996).

\bibitem{L74}
J. F. Navarro, C. S. Frenk, and S. D. M. White, A Universal Density Profile from Hierarchical Clustering, Astrophys. J. 490, 493 (1997).

\bibitem{L75}
D. N. Spergel and P. J. Steinhardt, Observational Evidence for Self-Interacting Cold Dark Matter, Phys. Rev. Lett. {\bf84}, 3760 (2000).

\bibitem{L76}
W. Hu, R. Barkana, and A. Gruzinov, Fuzzy Cold Dark Matter: The Wave Properties of Ultralight Particles, Phys. Rev. Lett. {\bf85}, 1158 (2000).

\bibitem{L77}
L. Berezhiani and J. Khoury, Theory of dark matter superfluidity, Phys. Rev. D {\bf92}, 103510 (2015).

\bibitem{L78}
B. Carr and F. K\"{u}hnel, Primordial black holes as dark matter candidates, SciPost Phys. Lect. Notes {\bf48}, 1 (2022).

\bibitem{L79}
A. Escriv\`{a}, F. K\"{u}hnel, Y. Tada, Primordial Black Holes, arXiv:2211.05767.

\bibitem{ADO2}
Z. Yousaf, B. Almutairi, M. Rizwan, T. Ganesan,  M. Z. Bhatti, Modeling less complex fuzzy dark matter cylindrical wormholes,  Int. J. Mod. Phys. D  {\bf34}, 2450072  (2025).

\bibitem{ADO6}
Z. Yousaf, M. Rizwan, M. Alshammari, O. Abdullah Almatroud, S. Alshammari, M. Mossa Al-sawalha, Minimally complex fuzzy wormhole models in Einstein-Cartan theory of gravity, Phys. Dark Univ. {\bf50}, 102082  (2025).

\bibitem{ADO4}
L. M. Abdalgadir, S. Khan,  H. M. A. Mahmoud, Fuzzified Hayward black hole configurations within dark matter halos,  Int. J. Geom. Methods Mod. Phys. \textbf{26502324} (2026).

\bibitem{ADO3}
A. H. Alblowy, M. Rizwan, N. Iqbal, W. W. Mohammed, H. Asad, Testing complexity to diagnose wormholes existence: static and spherically symmetric case. Eur. Phys. J. C {\bf86}, 45 (2026).

\bibitem{ADO5}
M. Alshammari, M. Rizwan, O. Abdullah Almatroud, M. Z. Bhatti, S. Alshammari, Z. Yousaf, Imprints of holographic dark energy and minimally deformed wormholes in general relativity, Phys. Dark Univ. {\bf51}, 102219
(2026).

\bibitem{ADO1}
Z. Yousaf, Viscous fluid cosmology: A path to cosmic acceleration, Phys. Dark Univ. {\bf 48}, 
101884 (2025).

\bibitem{L80}
V. V. Kiselev, Vector field and rotational curves in dark galactic halos, Class. Quantum Grav. {\bf22}, 541 (2005).

\bibitem{L81}
M. H. Li and K. C. Yang, Galactic dark matter in the phantom field, Phys. Rev. D {\bf86}, 123015 (2012).

\bibitem{L88}
Z. Xu, J. Wang, and X. Hou, Kerr-anti-de Sitter/de Sitter black hole in perfect fluid dark matter background, Class. Quantum Grav. {\bf35}, 115003 (2018).

\bibitem{L89}
S. Haroon, M. Jamil, K. Jusufi, K. Lin, and R. B. Mann, Shadow and Deflection Angle of Rotating Black Holes in Perfect Fluid Dark Matter with a Cosmological Constant, Phys. Rev. D {\bf99}, 044015 (2019).

\bibitem{L82}
X. Hou, Z. Xu, and J. Wang, Rotating black hole shadow in perfect fluid dark matter, J. Cosmol. Astropart. Phys. {\bf12}, 040 (2018).

\bibitem{L83}
H. X. Zhang, Y. Chen, T. C. Ma, P. Z. He, and J.B. Deng, Bardeen black hole surrounded by perfect fluid dark matter, Chin. Phys. C {\bf45}, 055103 (2021).

\bibitem{L84}
J. Li and C. Jiang, Particles collision near rotating black hole in perfect fluid dark matter, Eur. Phys. J. Plus {\bf137}, 1142 (2022).

\bibitem{L85}
D. Pugliese and Z. Stuchl\'{\i}k, Dark matter effect on black hole accretion disks, Phys. Rev. D {\bf106}, 124034 (2022).

\bibitem{L86}
K. Jusufi, Quasinormal modes of black holes surrounded by dark matter and their connection with the shadow radius, Phys. Rev. D {\bf101}, 084055 (2020).

\bibitem{VisserPF}
M. Visser, The Kiselev black hole is neither perfect fluid, nor is it quintessence, Class. Quantum Grav. {\bf37}, 045001 (2020).

\bibitem{Potapov2016}
A. A. Potapov, G. M. Garipova, and K. K. Nandi, Revisiting perfect fluid dark matter: Observational constraints from our galaxy, Phys. Lett. B {\bf753}, 140--146 (2016), \href{https://doi.org/10.1016/j.physletb.2015.11.057}{doi:10.1016/j.physletb.2015.11.057}.

\bibitem{Kumar2019}
S. Kumar, R. C. Nunes, and S. K. Yadav, Testing the warmness of dark matter, Mon. Not. R. Astron. Soc. {\bf490}, 1406--1414 (2019), \href{https://doi.org/10.1093/mnras/stz2676}{doi:10.1093/mnras/stz2676}.

\bibitem{L90}
S. M. A. S. Bukhari and L. G. Wang, Seeing dark matter via acceleration radiation, Phys. Rev. D {\bf109}, 045009 (2024).

\bibitem{L91}
R. M. Wald, Black hole entropy is the Noether charge, Phys. Rev. D {\bf48}, R3427(R) (1993).

\bibitem{L92}
S. Chandrasekhar, The Solution of Dirac's Equation in Kerr Geometry, Proc. Roy. Soc. Lond. A {\bf349}, 571 (1976).

\bibitem{L93}
S. Chandrasekhar, The Mathematical Theory of Black Holes, Fundam. Theor. Phys. {\bf9}, 5 (1984).

\bibitem{L94}
Z. Zhao and Y. X. Gui, The Connection between Unruh scheme and Damour-Ruffini scheme in Rindler space-time and $\eta-\varepsilon$ space-time, Nuovo Cim. B {\bf109}, 355 (1994).

\bibitem{L95}
T. Damour and R. Ruffini, Black Hole Evaporation in the Klein-Sauter-Heisenberg-Euler Formalism, Phys. Rev. D {\bf14}, 332 (1976).

\bibitem{L96}
S. M. Barnett and P. M. Radmore, \textit{Methods in Theoretical Quantum Optics}, Oxford University Press, Oxford (2002).

\bibitem{L97}
T. Baumgratz, M. Cramer, and M. B. Plenio, Quantifying coherence, Phys. Rev. Lett. {\bf113}, 140401 (2014).

\bibitem{L98}
M. B. Plenio, Logarithmic Negativity: A Full Entanglement Monotone That is not Convex, Phys. Rev. Lett. {\bf95}, 090503 (2005).
\end{thebibliography}
\end{document}